\documentclass[journal]{IEEEtran}

\usepackage{graphicx}
\usepackage{amsmath}
\usepackage{hhline}
\usepackage{stfloats}
\usepackage{hyperref}
\usepackage[export]{adjustbox}

\usepackage{schemabloc}

\usepackage{tabularx}

\usepackage{caption}
\usepackage{subcaption}
\usepackage[pscoord]{eso-pic}

\newcommand{\placetextbox}[3]{
  \setbox0=\hbox{#3}
  \AddToShipoutPictureFG*{
    \put(\LenToUnit{#1\paperwidth},\LenToUnit{#2\paperheight}){\vtop{{\null}\makebox[0pt][c]{#3}}}%
  }%
}%

\begin{document}
\title{Development of Dual-Gain SiPM Boards for Extending the Energy Dynamic Range}

\author{Daniel~Shy,
        Richard~S.~Woolf,
        Eric~A.~Wulf,
        Clio~C.~Sleator,
        Mary~Johnson-Rambert,
        W.~Neil~Johnson,
        J.~Eric~Grove,
        and~Bernard~F.~Phlips
        \thanks{This work was sponsored by NASA-APRA (NNH18ZDA001N-APRA). D. Shy is supported by the U.S. Naval Research Laboratory's Jerome and Isabella Karle Fellowship.}
        \thanks{D. Shy, R. Woolf, E. Wulf, C. Sleator, M. Johnson-Rambert, J. E. Grove, and B. Phlips are with the Space Science Division, U.S. Naval Research Laboratory, 4555 Overlook Ave., SW, Washington, DC, 20375, United States of America}
        \thanks{W. N. Johnson is with the Technology Service Corporation, Arlington, VA, 22202, United States of America}
        }


\maketitle

\begin{abstract}

Astronomical observations with gamma rays in the range of several hundred keV to hundreds of MeV currently represent the least explored energy range. To address this so-called MeV gap, we designed and built a prototype CsI:Tl calorimeter instrument using a commercial off-the-shelf (COTS) SiPMs and front-ends which may serve as a subsystem for a larger gamma-ray mission concept. During development, we observed significant non-linearity in the energy response. Additionally, using the COTS readout, the calorimeter could not cover the four orders of magnitude in energy range required for the telescope. We, therefore, developed dual-gain silicon photomultiplier (SiPM) boards that make use of two SiPM species that are read out separately to increase the dynamic energy range of the readout. In this work, we investigate the SiPM's response with regards to active area ($3\times3 \ \mathrm{mm}^2$ and $1 \times 1 \ \mathrm{mm}^2$) and various microcell sizes ($10$, $20$, and $35 \ \mu \mathrm{m}$). We read out $3\times3\times6 \ \mathrm{cm}^3$ CsI:Tl chunks using dual-gain SiPMs that utilize $35 \ \mu \mathrm{m}$ microcells for both SiPM species and demonstrate the concept when tested with high-energy gamma-ray and proton beams. We also studied the response of $17 \times 17 \times 100 \ \mathrm{mm}^3$ CsI bars to high-energy protons. With the COTS readout, we demonstrate a sensitivity to $60 \ \mathrm{MeV}$ protons with the two SiPM species overlapping at a range of around $2.5-30 \ \mathrm{MeV}$. This development aims to demonstrate the concept for future scintillator-based high-energy calorimeters with applications in gamma-ray astrophysics.

\end{abstract}


\section{Introduction}
\label{sec:intro}
\placetextbox{0.5}{0.05}{\large\textsf{Distribution Statement A: Approved for public release. Distribution is unlimited.}}%

\IEEEPARstart{T}{he} region between several hundred keV to several hundred MeV presents one of the most under-explored regions of the astronomical sky~\cite{mcenery2019allsky}. The last major observatory dedicated to the MeV range was the Compton Gamma-Ray Observatory which was de-orbited in the year 2000~\cite{COMPTEL}. The difficulty associated with MeV gamma-ray instrumentation is largely due to physical phenomena in the MeV region that include low interaction cross sections, high background, and activation of the instrument. Nevertheless, observation in the MeV band will aid in the study of nucleosynthesis in supernova~\cite{nuclearSyn}, supermassive blackholes~\cite{paliya2019supermassive}, and aid in the multi-messenger observation of astrophysical objects~\cite{burns2019opportunities}.

\begin{figure}[h!]
  \centering
  \includegraphics[trim={0cm 0cm 0cm 0cm}, clip, width=0.75\linewidth]{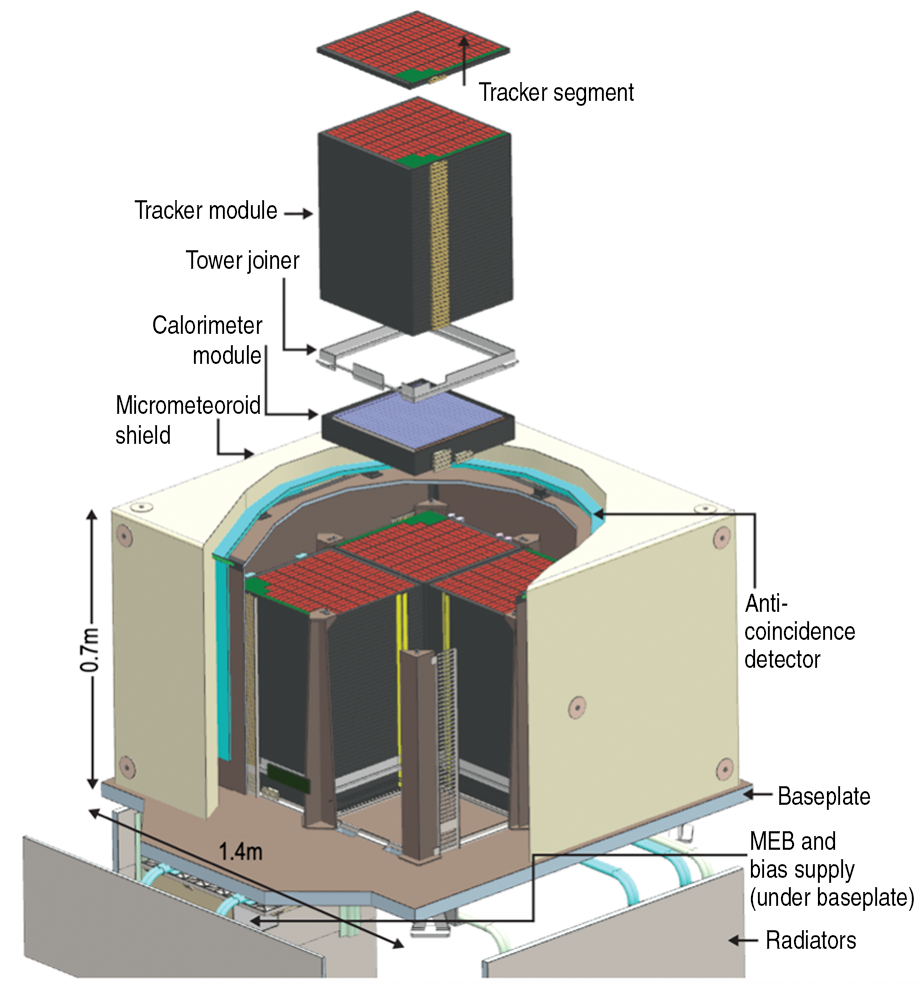}
  \caption{Exploded CAD view of the AMEGO-X gamma-ray telescope. There are 4 `towers' within the AMEGO-X instruments. Each tower consists of a tracker on top and a CsI-based calorimeter below. An anti-coincidence detector surrounds the entire instrument. Figure is adapted from~\cite{amegox}.}
  \label{fig:amegox}
\end{figure}

\begin{figure}[h!]
  \centering
  \includegraphics[trim={0cm 0cm 0cm 0cm}, clip, width=0.7\linewidth]{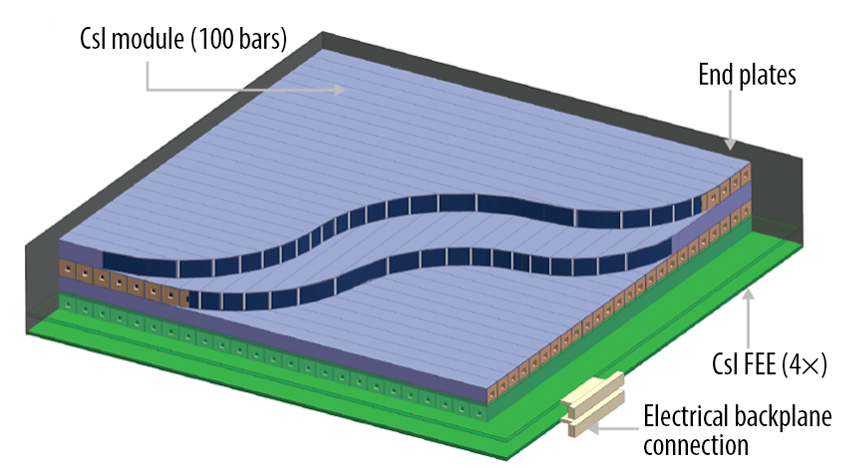}
  \caption{Cutaway view of a single CsI Calorimeter in a tower onboard AMEGO-X. This shows 4 layers of the calorimeter for a single tower with the bars organized in a hodoscopic fashion. Figure is adapted from~\cite{amegox}.}
  \label{fig:amegoxCalorimeter}
\end{figure}

In response, a medium-class Explorers (MIDEX) mission concept named the All-sky Medium-Energy Gamma-ray Observatory-eXplorer (AMEGO-X) is proposed to tackle this observational gap. AMEGO-X will act as a Compton and pair telescope~\cite{amegox}. The telescope consists of two subsystems, a tracker and a calorimeter, and are designed to assemble as four identical towers arranged in a $2\times2$ array. The tracker, led by NASA's Goddard Space Flight Center, is composed of layers of silicon complementary metal-oxide-semiconductor (CMOS) monolithic active pixel sensors~\cite{astroPix}. It will record the position and energy deposited during a scattering event or track the particles from any ensuing electromagnetic showers. The other subsystem is a thallium-doped CsI (CsI:Tl) calorimeter in development by the U.S. Naval Research Laboratory (NRL). The calorimeter will act as the absorber plane for scattered events and catch the ensuing electromagnetic showers from high-energy gamma rays. Fig.~\ref{fig:amegox} shows a computer-aided design (CAD) drawing of the gamma-ray telescope on board AMEGO-X. The figure also presents an exploded view of a single tower. Fig.~\ref{fig:amegoxCalorimeter} shows the CAD model of the CsI calorimeter for a single tower. Its design is hodoscopic in nature such that each layer contains rows of CsI logs. The direction of the logs alternates orthogonally for each layer.  Currently, the logs are slated to be $15 \times 15 \times 388 \ \mathrm{mm}^3$ in volume with the calorimeter consisting of 24 logs per layer for a total of four layers.

To improve the technological readiness level (TRL) of the calorimeter, we developed a CsI calorimeter R\&D prototype for the ComPair balloon-borne telescope~\cite{compair}, which utilizes $16.7 \times 16.7 \times 100 \ \mathrm{mm}^3$ CsI:Tl logs read out by a $2\times2$ array of $6 \times 6 \ \mathrm{mm}^2 $ SiPMs~\cite{woolf2019development}. The SiPMs are commercial off-the-shelf (COTS) arrays with a specific model ARRAYJ-60035-4P-PCB by onsemi~\cite{jSeriesArray}. The prototype calorimeter, which adopts the design aspects of the \textit{Fermi} Large Area Telescope's (LAT) calorimeter, is modular and consists of five layers of CsI:Tl logs arranged in rows of six. The logs are also arranged in a hodoscopic fashion. Read out on each end with silicon photomultipliers (SiPMs)~\cite{siPMs} allows for the reconstruction of energy and depth for each interaction. SiPMs present a great many advantages for reading out scintillators when compared to photomultiplier tubes (PMT). Among them are their low power and voltage requirements, insusceptibility to magnetic fields, low weight, and compact form. However, the number of microcells limits their signal in addition to affecting their susceptibility to radiation damage~\cite{MITCHELL2022167163}.

\begin{figure}[h!]
  \centering
  \includegraphics[trim={0cm 0cm 0cm 0cm}, clip, width=1\linewidth]{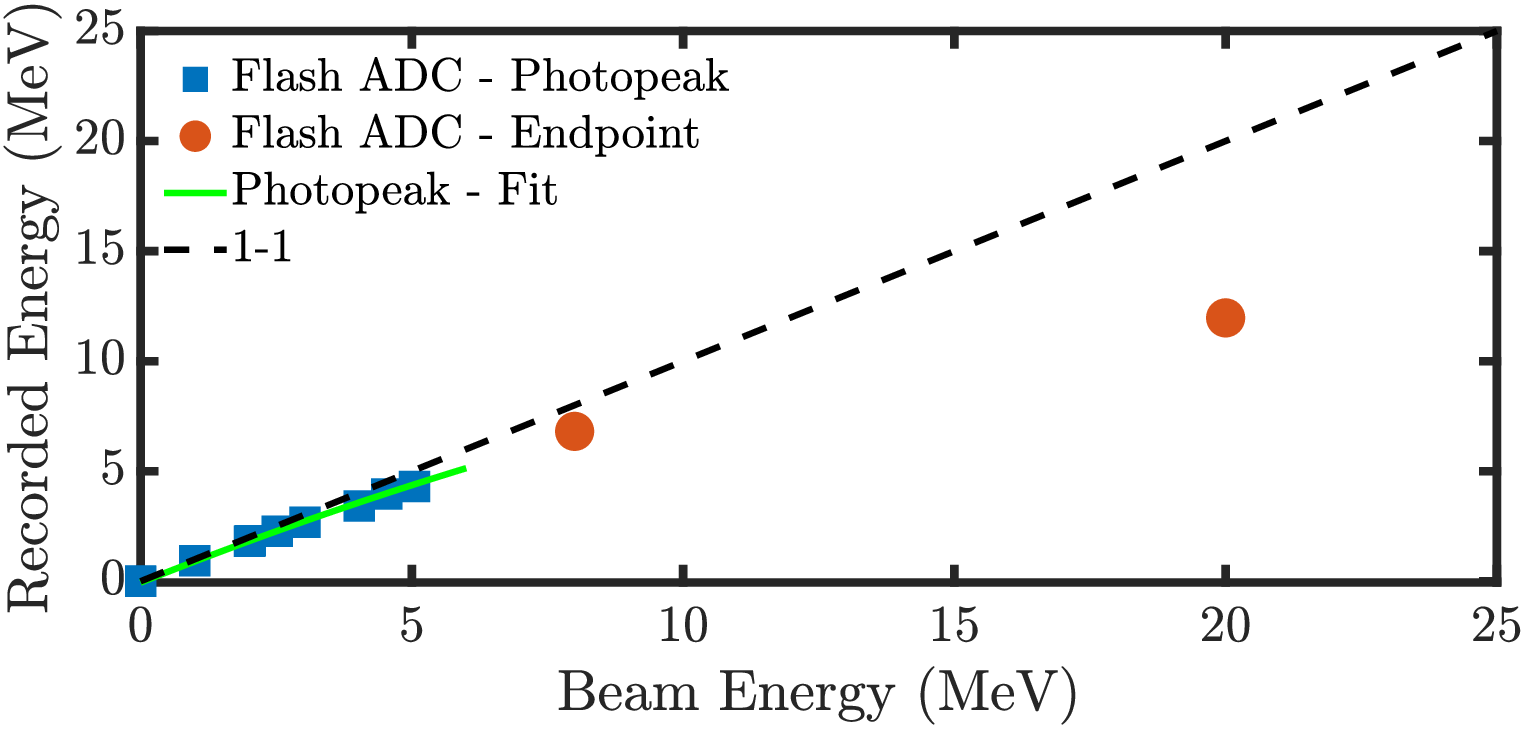}
  \caption{The scatter points present the recorded energy versus the beam energy and display the non-linear response. The recorded energy value represents the energy reconstructed using a calibration created using a 1.02 MeV line (from the 2.02 MeV double-escape peak~\cite{knoll2010radiation}). The dashed black line represents a 1-to-1 linear relationship. The solid green line represents a quadratic fit.}
  \label{fig:nonLinearity}
\end{figure}

At the Triangle University Nuclear Laboratory's (TUNL) High Intensity Gamma-ray Source (HIGS) facility~\cite{higs}, we irradiated the prototype calorimeter system with mono-energetic gamma rays that vary from $2-30 \ \mathrm{MeV}$~\cite{woolf2019development}. Fig.~\ref{fig:nonLinearity} presents the recorded energy calibrated using a $1 \ \mathrm{MeV}$ line vs. the known beam energy. The plot reveals the non-linearity in the system which is possibly due to the readout electronics~\cite{SiPHRA}, scintillator light-yield non-proportionality~\cite{knoll2010radiation}, and SiPM non-linearity. The SiPM non-linearity may arise from the microcell size as well as the full active area. As reported in~\cite{siPMNonlinearity, SiPMNonlinearityOG}, the non-linearity can be introduced by the limited number of pixels and active area. The performance of the prototype does not meet our desired upper energy range of $400 \ \mathrm{MeV}$. Nevertheless, in principle, such a response can be fitted and corrected but will naturally limit the dynamic range of the system~\cite{Regazzoni_2017}.

In response, we developed custom dual-gain SiPMs that utilize two species of SiPMs, one optimized for low-energy ranges and one for high-energy events. The two species have different-sized active areas and might contain a mix of microcell sizes. The concept behind multi-sized SiPMs is inspired by the \textit{Fermi} LAT CsI calorimeter readout~\cite{groveCalorimeter} that consists of dual-range PIN diodes to cover the $5 \ \mathrm{MeV}- 60 \ \mathrm{GeV}$ range for a single crystal. Although this work is sponsored to develop readout in support of AMEGO-X, dual-gain SiPMs could be applied to any scintillator-based calorimeters such as ones proposed for use on GECCO~\cite{GECCO, geccoSPIE} or ASTROGAM~\cite{astrogam}.  This work reports on the development and testing of these custom SiPM to read out CsI:Tl scintillators. Sec.~\ref{sec:dualgain} introduces the custom SiPMs. Sec.~\ref{sec:higs} presents their response to gamma rays while Sec.~\ref{sec:ucdavis} reports on the SiPM's response to high-energy protons. Sec.~\ref{sec:Discussion} discusses the results and shows the spectral unification of the two SiPMs.

\section{Custom Optical Sensor Development}
\label{sec:dualgain}

\begin{figure}[h!]
  \centering
  \includegraphics[trim={0cm 0cm 0cm 0cm}, clip, width=1\linewidth]{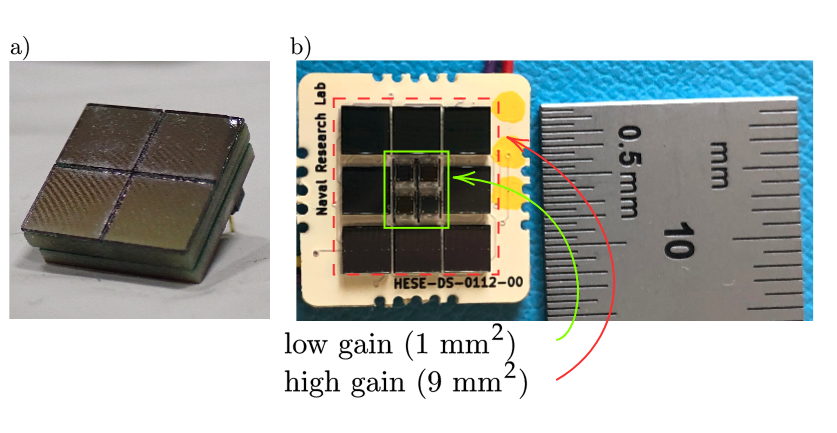}

  \caption{(a) Onsemi Quad-SiPM design utilized in the ComPair CsI calorimeter prototype. (b) The NRL custom design dual-gain SiPM with the low-gain group outlined in solid green while the high-gain group is outlined in dashed red.}
  \label{fig:dualGainSiPM}
\end{figure}

\begin{figure}[h!]
  \centering
  \includegraphics[trim={0cm 0cm 0cm 0cm}, clip, width=1\linewidth]{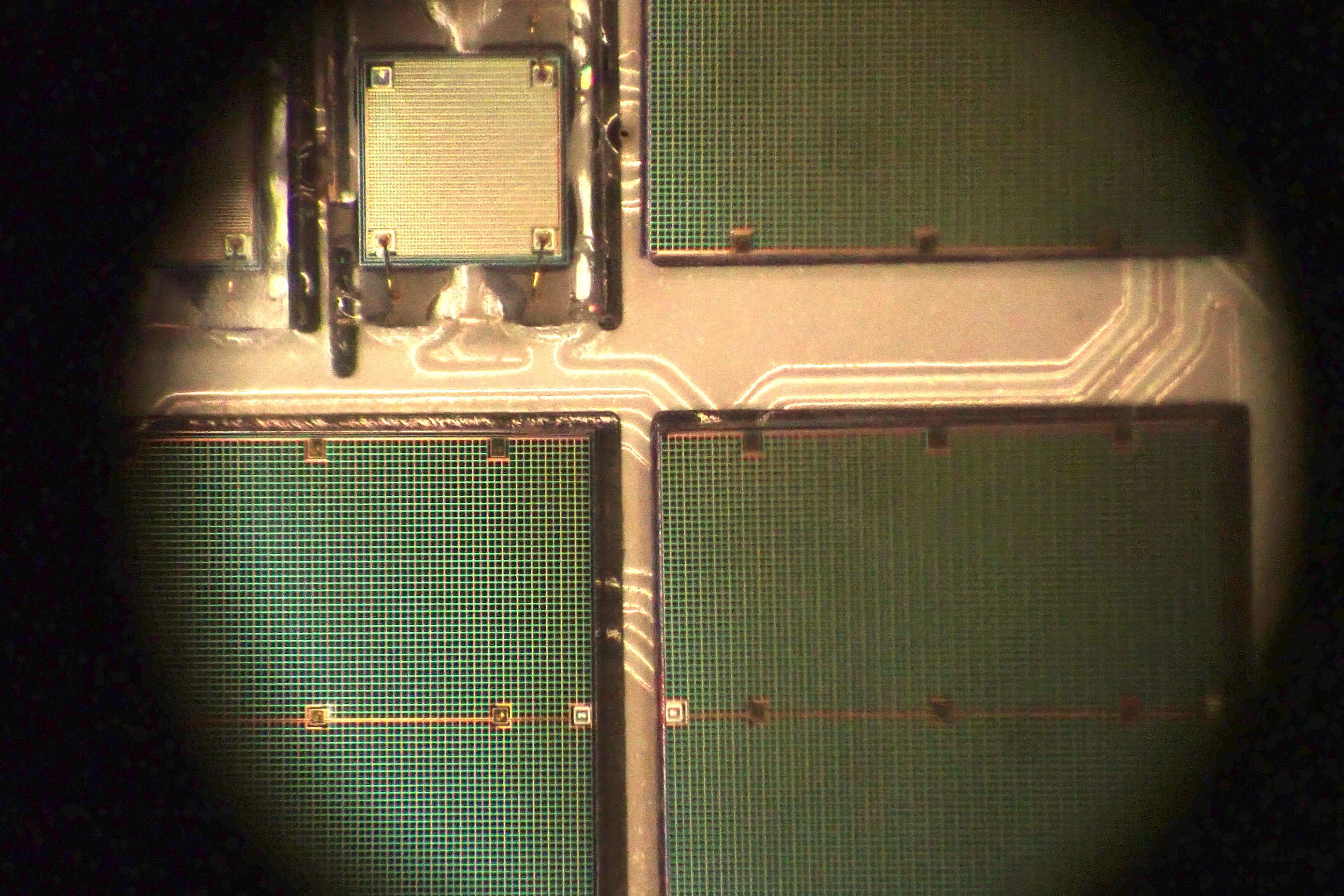}
  \caption{Microscopic view of the dual gain SiPM boards. The image zooms in on one $1 \ \mathrm{mm}^2$ SiPM (shown in the upper-left quadrant) while three $9 \ \mathrm{mm}^2$ SiPMs are shown in the remaining quadrants.
}
  \label{fig:microDualGain}
\end{figure}

The dual-gain SiPMs concept relies on fielding two different SiPM species with differently sized active areas and microcells. This concept aims to mitigate the saturation of the SiPMs and electronics.   Fig.~\ref{fig:dualGainSiPM}a shows the SiPMs utilized in the ComPair CsI calorimeter prototype~\cite{woolf2019development} while Fig.~\ref{fig:dualGainSiPM}b shows the layout of the custom board with two SiPM species, which we will refer to as `dual-gain SiPMs'. The dual-gain SiPMs consist of four $1\times1 \ \mathrm{mm}^2$ (`high-energy') SiPMs that are placed in the center of the carrier board to cover the high-energy range. They are surrounded by eight (`low-energy') SiPMs, each with an active area of $3 \times 3 \ \mathrm{mm}^2$ that will cover the low-energy range. Hereafter, the group of four $1\times1 \ \mathrm{mm}^2$ are referred to as as $1 \ \mathrm{mm}^2$ and eight $3 \times 3 \ \mathrm{mm}^2$ as $9 \ \mathrm{mm}^2$, based on their active area. Both SiPMs were manufactured by onsemi. Fig.~\ref{fig:microDualGain} shows a microscope view of the dual-gain SiPM.

\begin{figure}[h!]
  \centering
  \includegraphics[trim={0cm 0cm 0cm 0cm}, clip, width=1\linewidth]{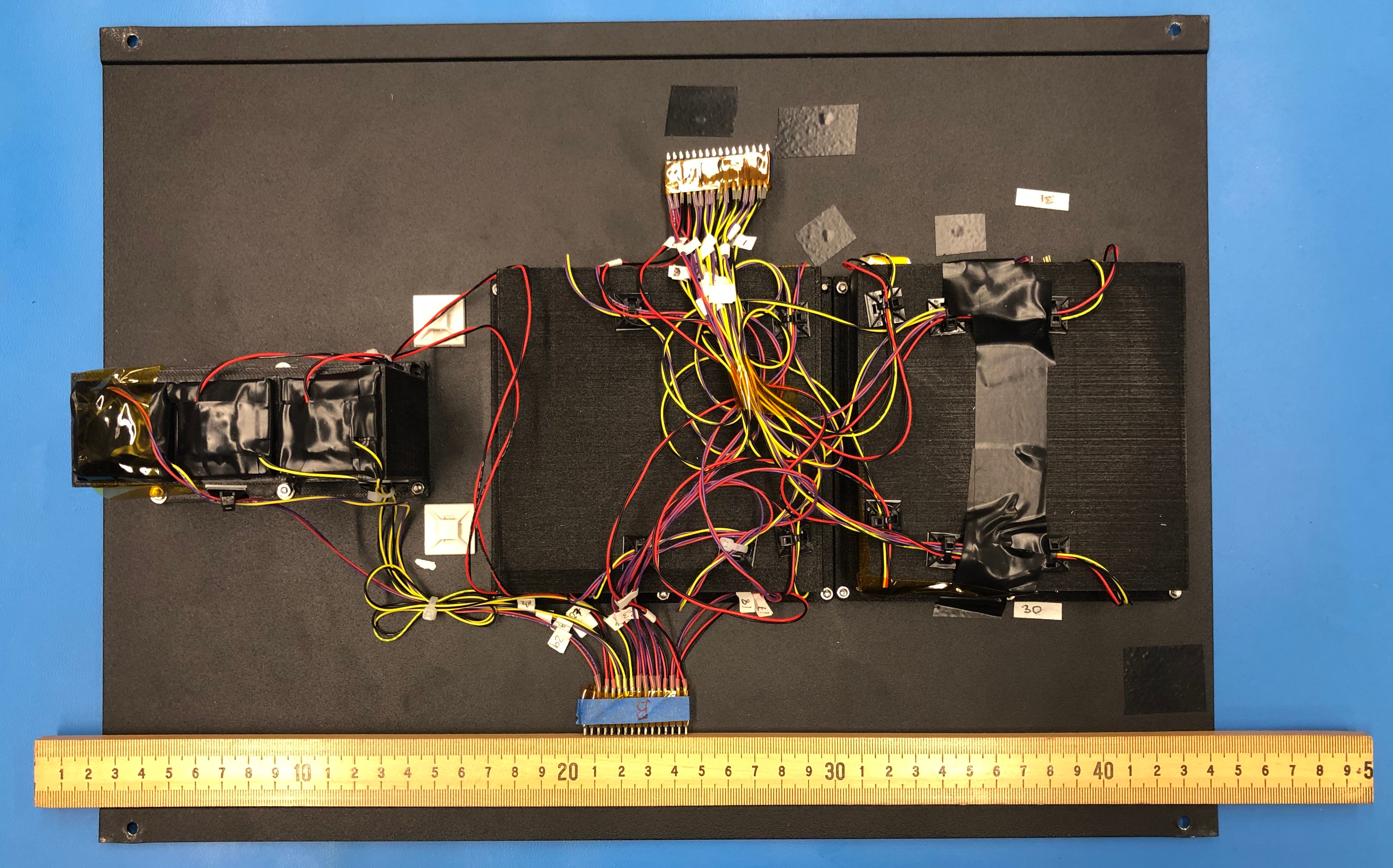}
  \caption{The scintillators utilized in the experiment were arranged in an array of 3 chunks (left) and 6 logs (right) that were attached to the lid of the experiment's box. The readout electronics would then be hooked up to each SiPM group via the connectors located above and below the array.}
  \label{fig:barSetup}
\end{figure}

We investigated the dual-gain design on $3\times 3 \times 6 \ \mathrm{cm}^3$ CsI chunks. We used the chunks due to their larger detection efficiency as well as serving as a potential geometry for an array-like calorimeter. We also studied CsI logs $(1.67\times 1.67 \times 10 \ \mathrm{cm}^3)$ that matched the current calorimeter prototype. While the logs utilize a SiPM on each end (the small faces), the chunks were read out on only one of the $3 \times 3 \ \mathrm{cm}^2$ side. Each scintillator crystal is wrapped with two layers of Tetratex and then warped with electrical tape to maintain light tightness. They were then placed in an additional box to maintain further light tightness. Fig.~\ref{fig:barSetup} shows the layout of the scintillators used in the experiment. Each scintillator was placed into a 3D-printed assembly. In the following experiments, the beam's vector was always out of the page.

Table~\ref{tab:versionSiPM} summarizes the different custom SiPM versions that were developed for initial testing. Note that these designs are used to demonstrate an initial concept rather than a slated design. For the experimental component of this work, we manufactured two logs and one chunk for a given SiPM version. Table~\ref{tab:SiPM_name} summarizes the part number of the utilized SiPMs.

\begin{table}[h!]
  \centering
  \caption{Characteristics of the explored dual-gain SiPMs with a description of each gain group. We define a SiPM group as a set of sensors of the same species.}
	\begin{tabularx}{\linewidth}{l|X|X}
	    \hline
		Version & High-Energy, $1 \ \mathrm{mm}^2$ (group of 4) & Low-Energy, $9 \ \mathrm{mm}^2$ (group of 8) \\ \hline
		V1      & $10 \ \mu \mathrm{m}$ microcells         & $35 \ \mu \mathrm{m}$ microcells        \\
		V2      & $10 \ \mu \mathrm{m}$ microcells         & $20 \ \mu \mathrm{m}$ microcells    	 \\
		V3      & $35 \ \mu \mathrm{m}$ microcells         & $35 \ \mu \mathrm{m}$ microcells    	 \\
		\hline
	\end{tabularx}
	\label{tab:versionSiPM}
\end{table}

\begin{table}[h!]
  \centering
  \caption{Part numbers of the SiPMs utilized dual-gain SiPM boards.}
	\begin{tabularx}{\linewidth}{l|X|X}
	    \hline
		Version & High-Energy C-Series~\cite{cSeries} & Low-Energy J-Series~\cite{jSeries}           \\ \hline
		V1      & MICROFC-10010-SMT-TR1 &   MICROFJ-30035-TSV-TR         \\
		V2      & MICROFC-10010-SMT-TR1 &   MICROFJ-30020-TSV-TR    	 \\
		V3      & MICROFC-10035-SMT-TR1 &   MICROFJ-30035-TSV-TR     	 \\
		\hline
	\end{tabularx}
	\label{tab:SiPM_name}
\end{table}

\subsection{SiPM Signal Processing}
In the custom SiPMs, for a given SiPM group, all the anodes are summed together with a simple summing circuit while all the cathodes are set to ground. Each group is read out by an Ideas ROSSPAD channel~\cite{rosspad}. The ROSSPAD is composed of 4 SiPHRA ASICs~\cite{SiPHRA} where each channel has a 12-bit ADC. We utilized a 1600 $\mathrm{ns}$ shaping time, the longest available in the SiPHRA to match the long pulse length associated with CsI. We also set the input stage gain to the lowest possible setting to allow for the largest dynamic range. Finally, both SiPM groups were biased to $27.5 \ \mathrm{V}$.

Since the logs are read out on each end, we calculate the energy deposited with~\eqref{eq:E}, which is the square root of the product between the energy recorded by each SiPM group shown, also known as the geometric mean:

\begin{align}
\label{eq:E}
E &=\sqrt{E_1 \times E_2},
\end{align}

and the depth along the log as

\begin{align}
\label{eq:DOI}
\mathrm{depth} & = \frac{E_1-E_2}{E_1+E_2},
\end{align}

\noindent
where $E_1$ is the energy recorded by one side while $E_2$ represents the other side. As the chunks are read out by one side, we do not estimate the depth and the deposited energy is simply taken as the energy recorded by the SiPM.

\section{Gamma-ray Beam Experiments at the High Intensity Gamma-ray Source}
\label{sec:higs}

The dual-gain setup was tested at the HIGS facility. Due to beam time constraints, we only tested the chunks. The tested energies were $8, 15$ and $20 \ \mathrm{MeV}$. We also experimented with an $11 \ \mathrm{MeV}$ beam. However, during the run, we received unexpected results that were more consistent with a faulty beam. We therefore omit any analysis from the $11 \ \mathrm{MeV}$ beam. During the experiment, the high-energy SiPMs of V1 and V2 did not show any response and we therefore conclude the lower threshold on the $1 \ \mathrm{mm}^2$ SiPMs is higher than $20 \ \mathrm{MeV}$. However, V3 displayed an overlap between the two SiPM species. 

\begin{figure}[h!]
  \centering
  \includegraphics[trim={0cm 0cm 0cm 0cm}, clip, width=1\linewidth]{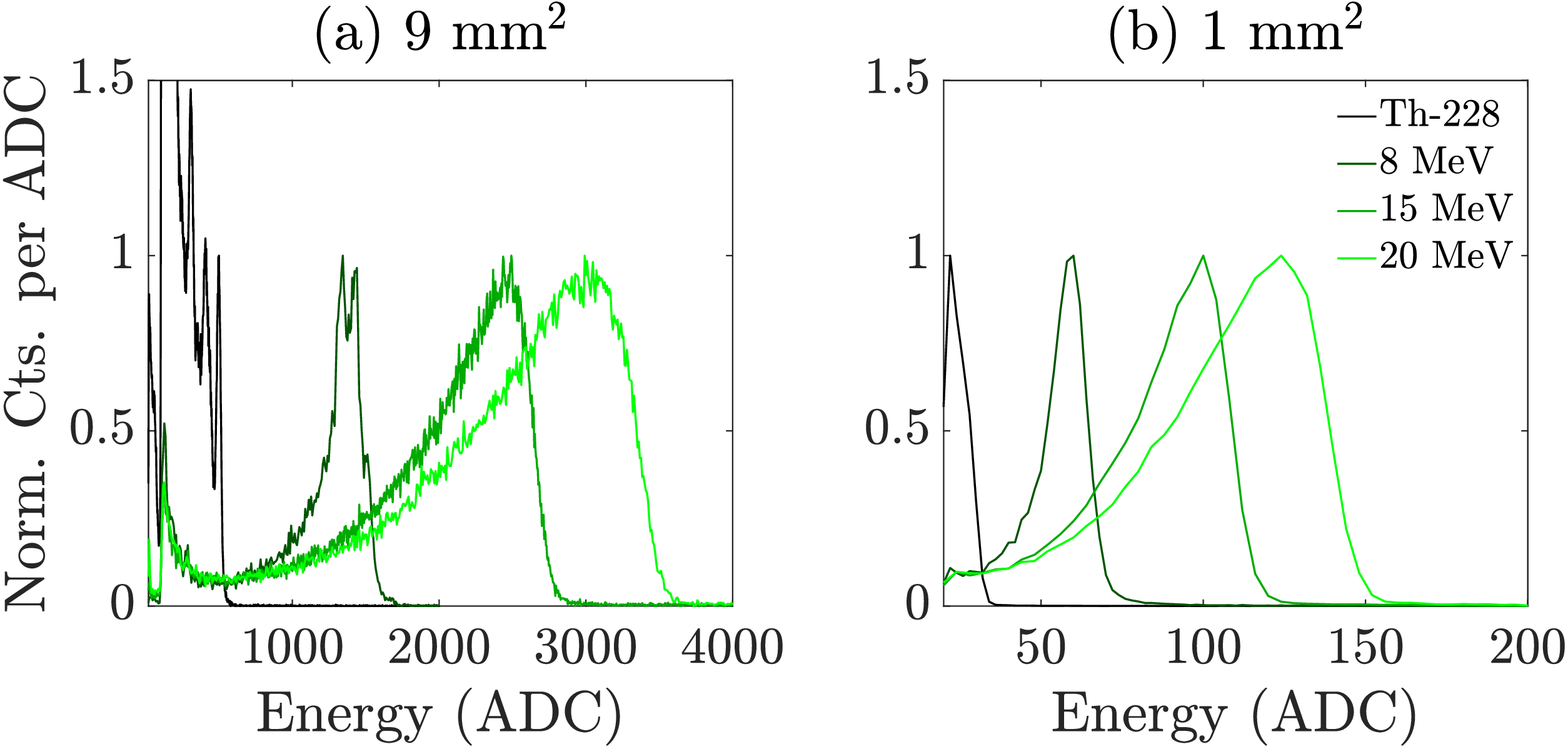}
  \caption{Spectral response of the (a) $9 \ \mathrm{mm}^2$ and (b) $1 \ \mathrm{mm}^2$ SiPMs on a $30\times 30 \times 60 \ \mathrm{mm}^3$ chunk scintillator to a Th-228 source and mono-energetic gamma rays. Note the different $x$-axis scales between (a) and (b).}
  \label{fig:HIGS_Spectra}
\end{figure}

Fig.~\ref{fig:HIGS_Spectra} plots the different spectra taken at the HIGS facility for the (a) $9 \ \mathrm{mm}^2$ and (b) $1 \ \mathrm{mm}^2$ of the V3 SiPMs. When using the $8 \ \mathrm{MeV}$ and Th-228 source, which emits a $2.614 \ \mathrm{MeV}$ gamma-ray, the full-energy peak is visible in the $9 \ \mathrm{mm}^2$ SiPMs along with the single and double escape peaks~\cite{knoll2000radiation}. At higher energies, due to the prevailing physics, those features have disappeared and the spectra have transformed into continua with their endpoint proportional to the maximum deposited energy of the gamma. When analyzing the $1 \ \mathrm{mm}^2$ SiPMs, we observe that they have a significantly lower ADC value for the same source energy when compared to the $9 \ \mathrm{mm}^2$. We also do not observe the full-energy peak as seen in the $9 \ \mathrm{mm}^2$ in the lower energies, but in general, the two groups show the same spectral trends at $8 \ \mathrm{MeV}$ and above. Appendix~\ref{sec:sims} discusses simulation efforts for this experiment showing similar trends as the experimental data.

\begin{figure}[h!]
  \centering
  \includegraphics[trim={0cm 0cm 0cm 0cm}, clip, width=1\linewidth]{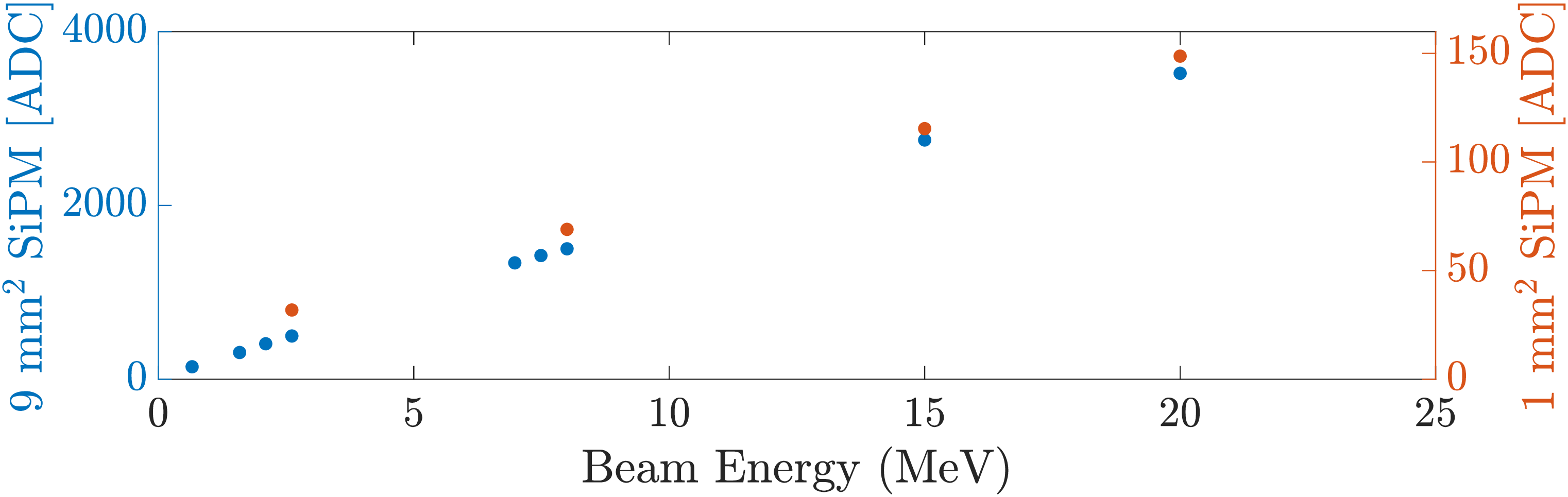}
  \caption{Peak locations for the $9 \ \mathrm{mm}^2$ and $1 \ \mathrm{mm}^2$ SiPMs for peaks associated with Th-228 source and mono-energetic gamma rays. Error bars for gamma-ray peak features reflect the full-width-at-half-maximum (FWHM) as calculated by a gaussian fit. Error bars for endpoint features reflect the $95\%$ confidence interval for the linear fit.}
  \label{fig:HIGS_Peaks}
\end{figure}

Fig.~\ref{fig:HIGS_Peaks} plots the recorded full-energy features as a function of the beam energy. When peaks are visible, a Gaussian fit was applied. When peaks were not visible (i.e. the 15 and 20 MeV runs), a linear fit was applied for the high-energy fall-off to find the endpoint energy. Since the beam flux was low enough to avoid chance coincidences, the highest recorded energy that is part of the spectral distribution most likely originates from a full-energy deposition of the beam gamma rays. We therefore apply a linear fit on the range of 75\%-25\% of the max peak on the high end of the spectrum. The plot also shows the clear overlap between the two species that ranges between $\sim2.6$ to $20  \ \mathrm{MeV}$ and linearity across the overlap region.

\section{High-Energy Proton Beam Experiments at the UC-Davis Crocker Nuclear Laboratory}
\label{sec:ucdavis}

\begin{figure}[h!]
  \centering
  \includegraphics[trim={0cm 2cm 0cm 2cm}, clip, width=1\linewidth]{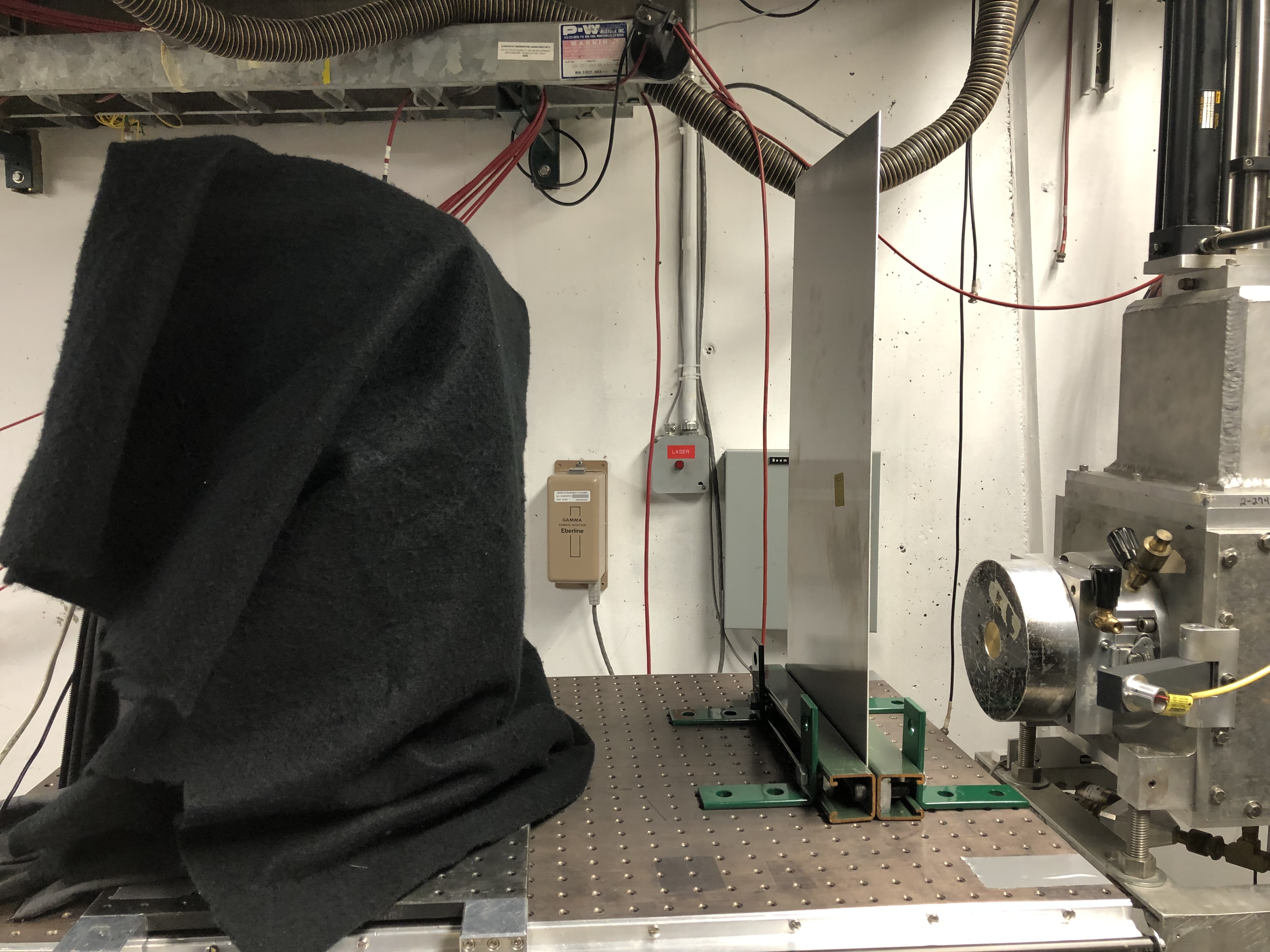}
  \caption{Experimental setup at the Crocker Nuclear Laboratory. The beam nozzle is located at the right of the image while the beam points towards the left. The detector box is placed on the left. The aluminum attenuator is placed in between the two. The instrument was covered with several layers of black cloth to maintain light tightness.
}
  \label{fig:experimentSetup}
\end{figure}

To study the high-energy performance of the custom SiPMs, we conducted high-energy proton beam experiments at the Crocker Nuclear Laboratory at the University of California$-$Davis~\cite{UCDavis}. The beam delivered $64 \ \mathrm{MeV}$ protons and was collimated using a $0.61 \ \mathrm{mm}$ diameter pinhole. We then added various aluminum attenuators throughout the experiment to reduce the beam's energy. Therefore, the beam energy ranged from $27$ to $60 \ \mathrm{MeV}$, which we estimated using SRIM~\cite{SRIM}. Fig.~\ref{fig:experimentSetup} shows the experimental setup containing both the logs and chunks.

\subsection{Experimental Results}

During the experiment, the $1 \ \mathrm{mm}^2$ SiPMs of V1 and V2 did not show any response to the high-energy protons and therefore conclude the $1 \ \mathrm{mm}^2$ SiPMs on those versions had a low threshold higher than $60 \ \mathrm{MeV}$ using the COTS ROSSPAD. Moreover, the $60 \ \mathrm{MeV}$ protons saturated the low-energy $9 \ \mathrm{mm}^2$ SiPMs which implies that there is no overlap in energy coverage between the two SiPM species; thus, we do not consider them candidates.

\begin{figure}[h!]
  \centering
  \includegraphics[trim={0cm 0cm 0cm 0cm}, clip, width=1\linewidth]{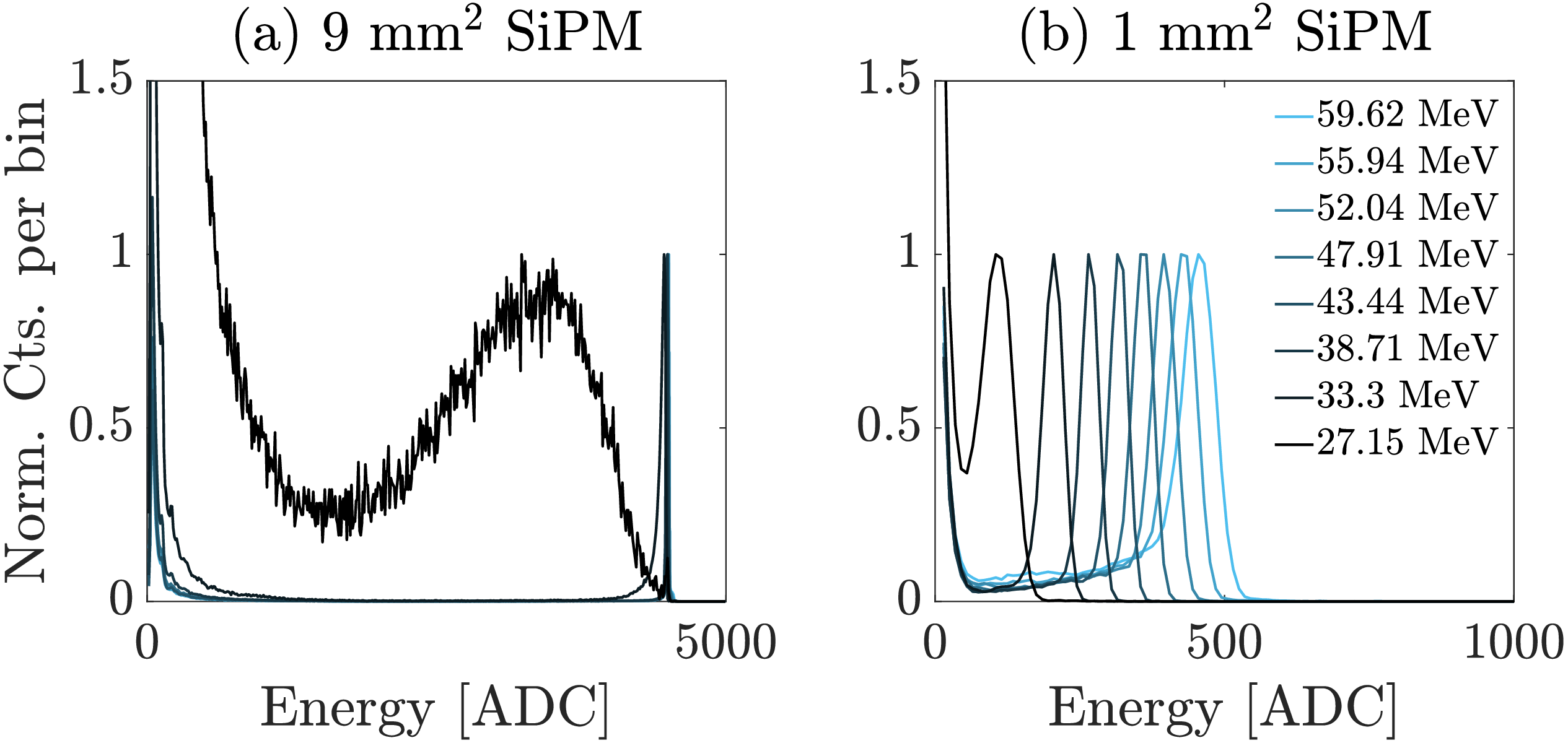}
  \caption{The response of a log to high-energy protons as measured by the V3 (a) $9 \ \mathrm{mm}^2$ and (b) $1 \ \mathrm{mm}^2$ SiPM. Note the overflow in (a).}
  \label{fig:A_V3_2_low__high_gain_spectra}
\end{figure}

The V3 prototype exhibited promising results in that there was an energy overlap between the two SiPM species. Fig.~\ref{fig:A_V3_2_low__high_gain_spectra} plots the responses to high-energy protons of (a) the $9 \ \mathrm{mm}^2$ and (b) $1 \ \mathrm{mm}^2$ SiPMs to reading a log.  Fig.~\ref{fig:A_V3_2_low__high_gain_spectra} shows that at $27 \ \mathrm{MeV}$, the $9 \ \mathrm{mm}^2$ SiPM is active showing a valid spectrum without overflows, while the $1 \ \mathrm{mm}^2$ SiPM also shows a similar response, although with much lower ADC values. As the beam energy increases, we observe the $9 \ \mathrm{mm}^2$ channel saturate while the $1 \ \mathrm{mm}^2$ SiPM group is sensitive to the increasing energy. Appendix~\ref{sec:additionalPlots} breakup Fig.~\ref{fig:A_V3_2_low__high_gain_spectra} based on energy to better display each energy response.

\begin{figure}[h!]
  \centering
  \includegraphics[trim={0cm 0cm 0cm 0cm}, clip, width=1\linewidth]{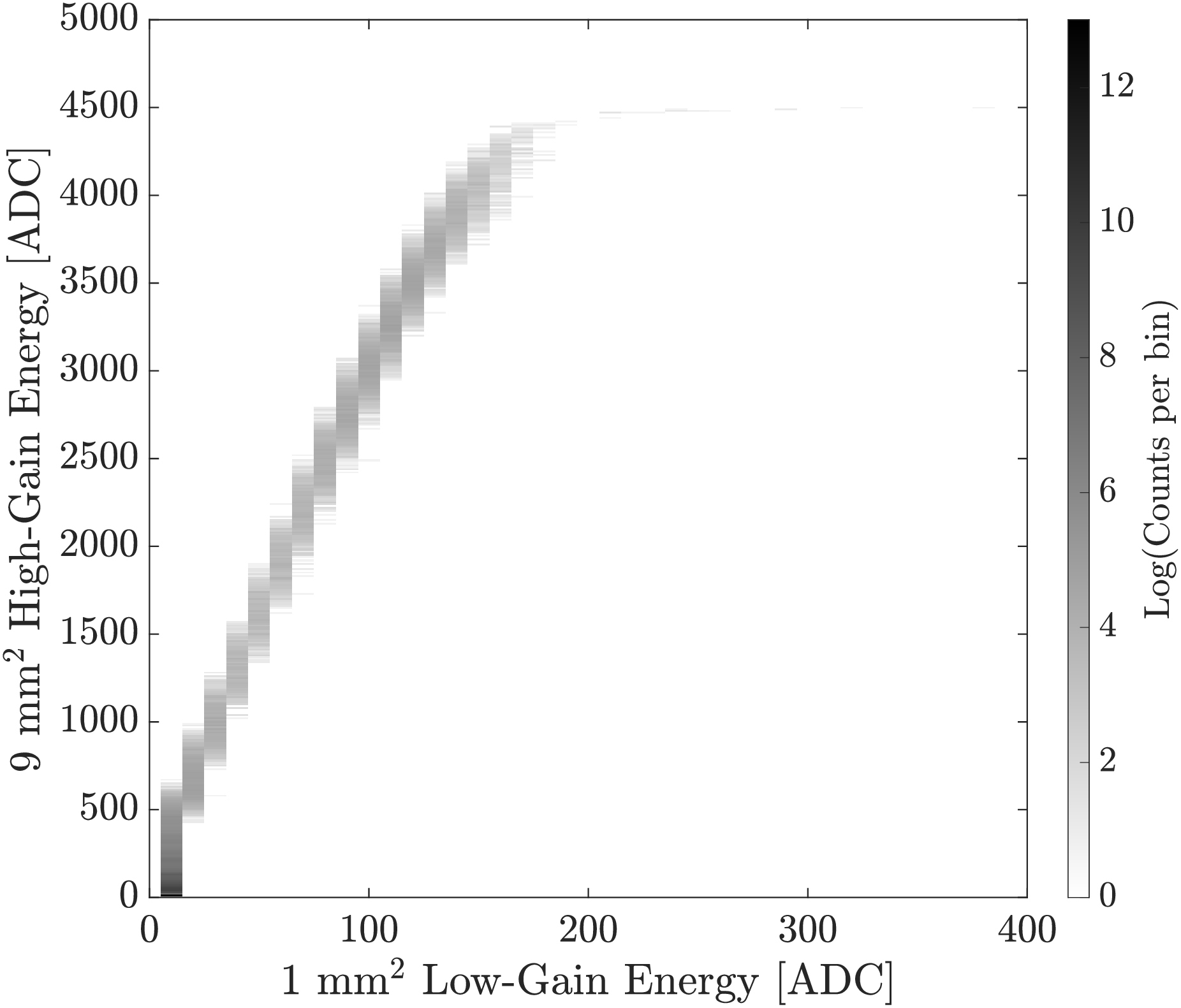}
  \caption{Bivariate responses of the low- and high-energy SiPMs to $27 \ \mathrm{MeV}$ protons. }
  \label{fig:bivariate_A_V3_2}
\end{figure}

The correlated nature of the two SiPM species is best shown in Fig.~\ref{fig:bivariate_A_V3_2}, which plots a bivariate histogram of the $1 \ \mathrm{mm}^2$ and $9 \ \mathrm{mm}^2$ SiPMs when reading out the logs. For low energies, there is a clear relationship between the two SiPMs up until where the $9 \ \mathrm{mm}^2$ SiPM saturates. Moreover, the $9 \ \mathrm{mm}^2$ SiPMs begin behaving non-linearly at around $3500$ ADCs.

\begin{figure}[h!]
  \centering
  \includegraphics[trim={0cm 0cm 0cm 0cm}, clip, width=1\linewidth]{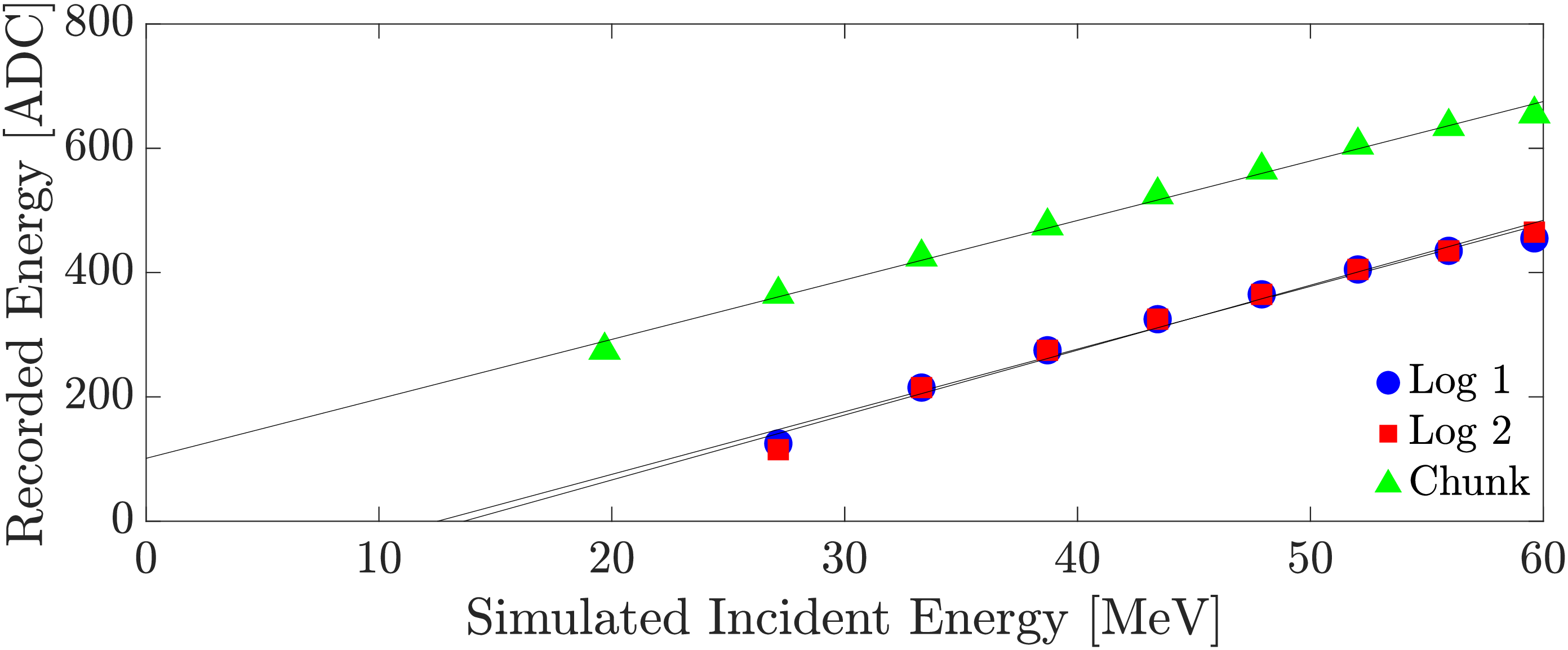}
  \caption{The responses of the $1 \ \mathrm{mm}^2$ SiPMs with the logs and chunks as a function of incident energy as simulated by SRIM. Error bars for the $x$-axis reflect the FWHM of the simulated beam energy incident on the scintillator. The $y$-axis error bars reflect the FWHM of the measured spectra.}
  \label{fig:spectraResponse}
\end{figure}

Fig.~\ref{fig:spectraResponse} plots the response of the $1 \ \mathrm{mm}^2$ SiPMs for the chunk and two logs. The error on the $y$-axis reflects the FWHM of the recorded spectrum while the error on the $x$-axis is the FWHM of the simulated spectrum. The addition of the error bars using the FWHM values is to give some indicate the resolution when compared to the energy blurring due to stochastic processes of the proton-matter interaction. Note that there are additional errors in the simulated energy that is not included. First, the simulation assumed pure aluminum for the attenuator, which the experiment did not utilize. In addition, any setup error that resulted in a possible slight tilt of the aluminum attenuators was not accounted for and implies that the simulated energies might be an overestimation due to the possible increased thickness.

\subsection{Depth of Interaction Along the Length of the Crystal Study}
\label{sec:doiStudy}

\begin{figure}[h!]
  \centering
  \includegraphics[trim={0cm 0cm 0cm 0cm}, clip, width=1\linewidth]{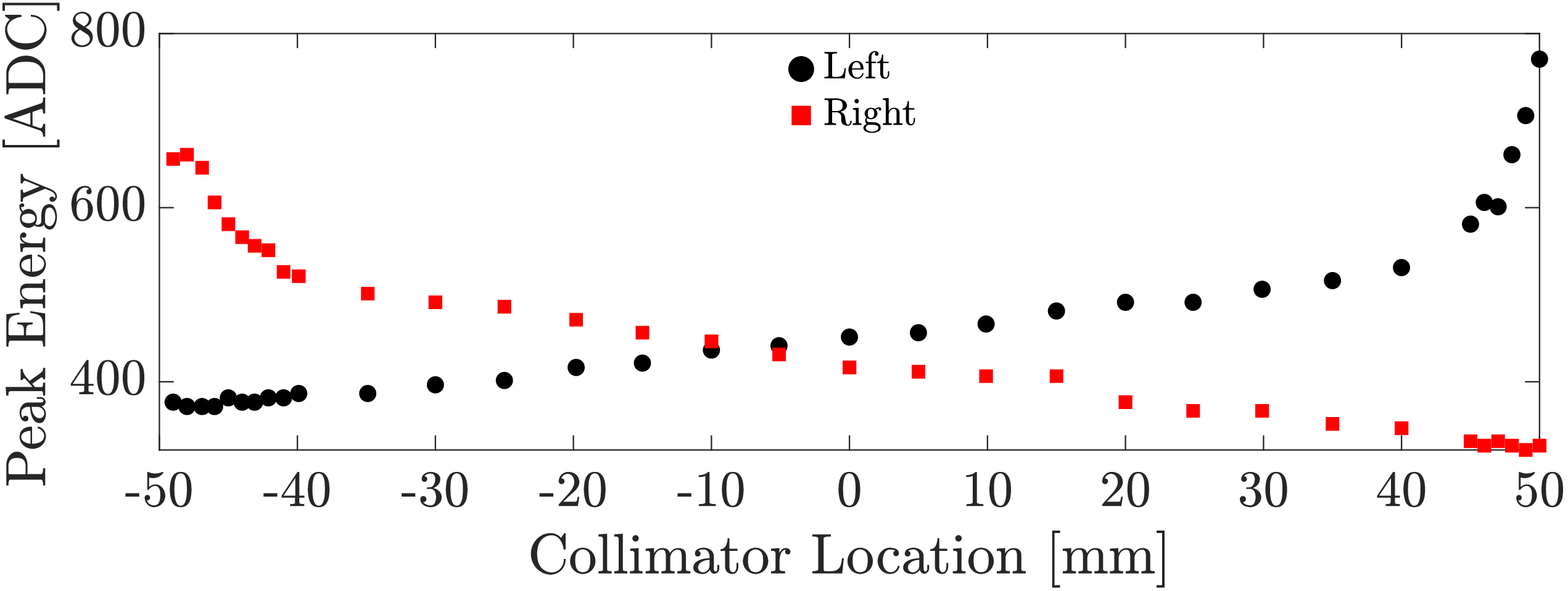}
  \caption{The responses of the left and right SiPM as a function of the collimator location.}
  \label{fig:DOIPeakSiPMResponse}
\end{figure}

We also performed a depth of interaction study where we scanned the length of the log with the proton pencil beam. We took $1 \ \mathrm{mm}$ steps for $1 \ \mathrm{cm}$ on each end of the log and took $5 \ \mathrm{mm}$ steps for the central $8 \ \mathrm{cm}$ of the $10 \ \mathrm{cm}$ bar. The setup was left unattenuated with aluminum, thus the estimated proton energy is $\sim 60 \ \mathrm{MeV}$. Fig.~\ref{fig:DOIPeakSiPMResponse} plots the responses of the left and right high-energy SiPMs as a function of the collimator location with $0 \ \mathrm{mm}$ representing the physical center of the log. In the range of about $-40$ to $40 \ \mathrm{mm}$, the SiPM's responses display a linear trend as a function of the interaction location relative to the SiPM. When the interaction occurs closer to the SiPM, the SiPM experiences the well-known `edge effect' where their light exposure deviates significantly from linearity when compared to the SiPM on the other end of the log~\cite{glastCalEngineering}.

\begin{figure}[h!]
  \centering
  \includegraphics[trim={0cm 0cm 0cm 0cm}, clip, width=1\linewidth]{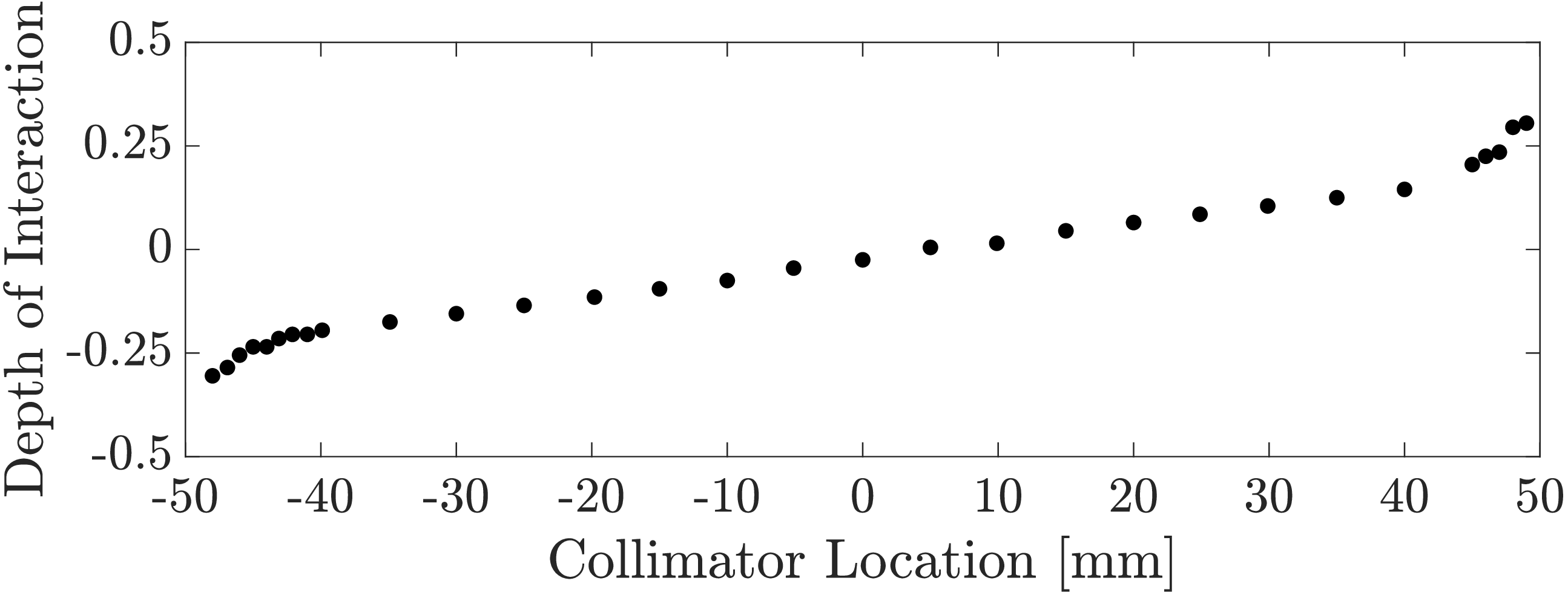}
  \caption{Calculated depth of interaction along the log as a function of collimator location.}
  \label{fig:doiLoc}
\end{figure}

\begin{figure}[h!]
  \centering
  \includegraphics[trim={0cm 0cm 0cm 0cm}, clip, width=1\linewidth]{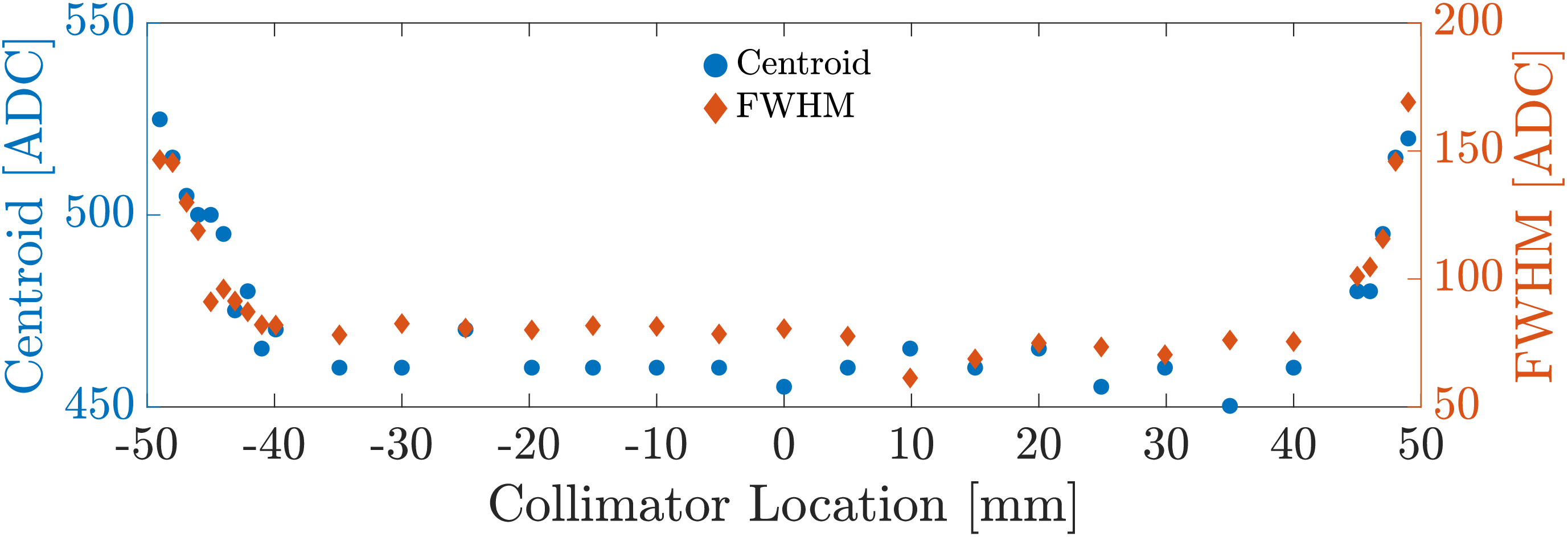}
  \caption{The peak centroid and resolution (FWHM) of the reconstructed full-energy peak as a function of collimator location. Note that the FWHM includes the resolution of the proton beam.}
  \label{fig:FWHMV_centroid_Loc}
\end{figure}

Fig.~\ref{fig:doiLoc} plots the calculated depth vs. the collimator position and shows a deviation from the general trend at the edges. Fig.~\ref{fig:FWHMV_centroid_Loc} also shows the edge effect on the peak centroid and resolution (FWHM). Not only does the centroid increase near the edges, but its resolution also degrades proportionally.

\section{Discussion}
\label{sec:Discussion}

This work demonstrates the dual-gain SiPMs concept. The most crucial observation was to check that the energy ranges of the two species overlapped, ensuring continuous coverage over the entire energy range of interest. The results from HIGS (Fig.~\ref{fig:HIGS_Peaks}) show the clear overlap in ranges with data points ranging from $2.6$ to $20 \ \mathrm{MeV}$ for the V3 SiPMs. It is in this region that we will declare the turnover point between the $9 \ \mathrm{mm}^2$ and $1 \ \mathrm{mm}^2$ SiPM group.

The goal of the experiment at the Crocker Nuclear Laboratory was to verify the overlap region and to study the high-energy response of the SiPMs. Protons were chosen as they can easily deposit more energy into the log than high-energy gammas. With the different proton energies, we were able to plot the response and estimate that the higher end of the energy range is around $400 \ \mathrm{MeV}$ for the logs and chunks alike when linearly extrapolating with the acquired data. The upper energy limit is derived using the ROSSPAD's 12-bit ADC range with the assumption that the SiPM/readout electronics maintain linearity throughout the energy range and that they do not saturate. In reality, both the SiPM and the ROSSPAD do experience non-linearity and therefore the true upper limit might be lower than $400 \ \mathrm{MeV}$. In addition, we did not have the opportunity to sample higher energies to study the non-linearity of the $1 \ \mathrm{mm}^2$ as well as its upper threshold. Therefore, with the aforementioned assumptions, the combination of the two SiPMs with the ROSSPAD may result in an energy range from $250 \ \mathrm{keV}$ to $400 \ \mathrm{MeV}$. 

We note that the linearity that is experienced by each group might differ, due to the limited number of pixels and the active area as is commonly described with~\eqref{eq:nonLinearity}:

\begin{align}
\label{eq:nonLinearity}
N_{\mathrm{cells \ fired}} = N_\mathrm{pixel} \left ( 1-e^{\frac{\epsilon N_\mathrm{photon} }{N_\mathrm{pixel}}} \right ),
\end{align}

where $N_{\mathrm{cells \ fired}}$ is the number of cells that fired and is proportional to the signal, $N_\mathrm{pixel}$ is the total number of pixels while $N_\mathrm{photon}$ is the total number of photons in a pulse. $\epsilon$ is the photon detection efficiency. However, this model does not include non-linear light yield from the scintillator, noise in the SiPM, and non-linearity in the readout electronics.

\begin{figure}[h!]
  \centering
  \includegraphics[trim={0cm 0cm 0cm 0cm}, clip, width=1\linewidth]{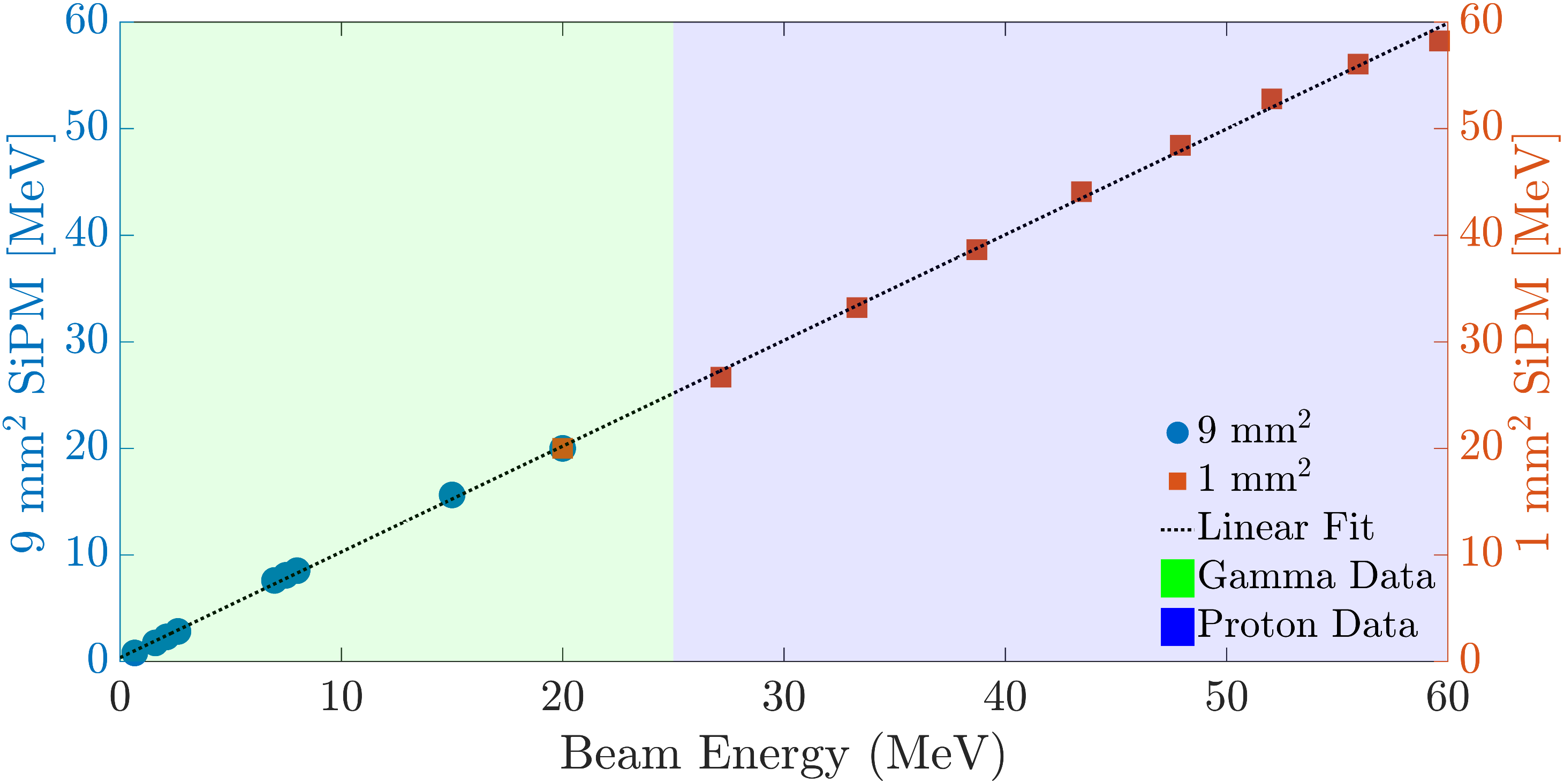}
  \caption{Unifying the $9 \ \mathrm{mm}^2$ and $1 \ \mathrm{mm}^2$ SiPM into a single plot utilizing both the gamma-ray and proton data. The error bars are similar to the ones utilized in Fig.~\ref{fig:HIGS_Peaks} and~\ref{fig:spectraResponse}. The shaded region signify when proton or gamma data is utilized.}
  \label{fig:unified_peaks}
\end{figure}

Fig.~\ref{fig:unified_peaks} plots unified data between the two species using both the gamma and proton data where we chose $20 \ \mathrm{MeV}$ as the transition energy. The energy calibration is derived from a linear correction applied to each SiPM group. This plot displays the dual-gain concept in which $9 \ \mathrm{mm}^2$ SiPMs are sensitive to ``low-energy'' events while the $1 \ \mathrm{mm}^2$ SiPMs take over for ``high-energy'' events.

During the log scanning effort in Sec.~\ref{sec:doiStudy}, we also irradiated the SiPMs with high-energy protons. In doing so, we have observed no overflow triggers in the $1 \ \mathrm{mm}^2$  SiPMs or events associated with proton-SiPM interactions. There, the protons entered along the plane of the SiPM. The negligible effect is most likely due to the small active region in the SiPM to which the proton deposits minimal energy. In addition, the SiPM is resistant to avalanche effects from one microcell to another in standard operations~\cite{sipmAvalanching}.

\section{Conclusion and Future Work}

This initial investigation demonstrates the responses of dual-gain SiPMs utilized to read out the scintillator. The low-energy SiPMs can cover the `low' energy range and the high-energy SiPMs will accommodate the `higher' energy range. While the high-energy SiPMs have a measured up to 60 MeV with protons data. According to our calculation, and assuming the SiPMs and readout system do not experience non-linearity, the dual-combination might provide a dynamic energy range from $250 \ \mathrm{keV}$ to $400 \ \mathrm{MeV}$ using the SiPHRA's 12-bit ADC. Dedicated tests on SiPMs response and photoelectron yields are needed to confirm this hypothesis. In the future, we plan to build a flight-like calorimeter that is intended to serve as a single tower for the AMEGO-X mission concept~\cite{amegox}. We are also investigating the usage of custom dual-gain SiPMs manufactured by Fondazione Bruno Kessler (FBK).

\section*{Acknowledgment}

The authors thank Alexander A. Moiseev for the opportunity to join his HIGS beam test. We are also grateful to Michael Backfish, the staff at the Crocker Nuclear Laboratory, and the accelerator physics group at High Intensity Gamma-ray Source (HIGS). We thank Ms. Emily Kong for simulation support in this project.

Much appreciation is given to the anonymous referees for the most valuable comments which have improved this manuscript.

\appendices

\section{Simulation of the System to High-Energy Gamma Rays}
\label{sec:sims}

\begin{figure}[h!]
  \centering
  \includegraphics[trim={0cm 0cm 0cm 0cm}, clip, width=1\linewidth]{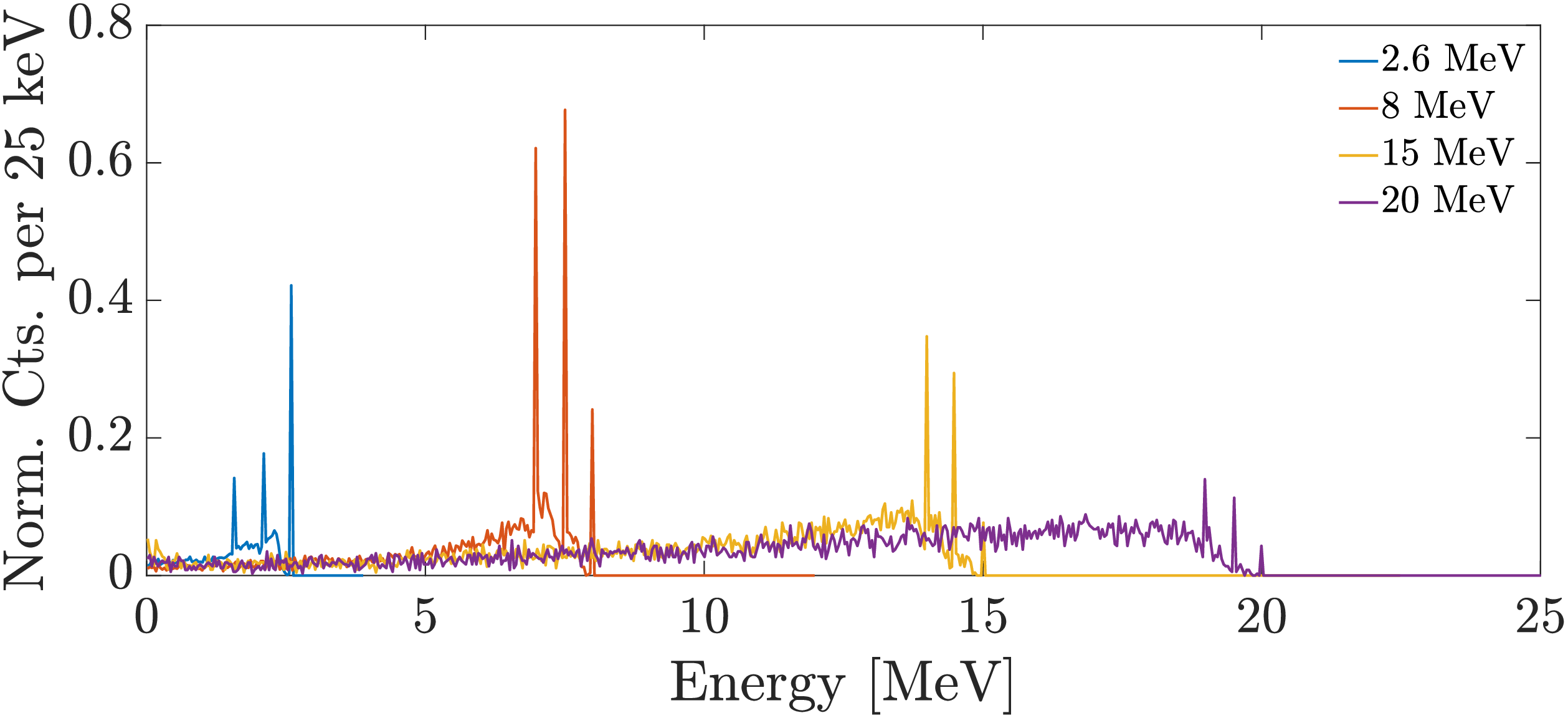}
  \caption{Simulated spectral responses of an ideal CsI chunk to several monoenergtic gamma rays.}
  \label{fig:gammaSimulations}
\end{figure}

This section presents simulations of the experimental setup of the CsI chunk using the SWORD simulation package~\cite{SWORD}. Like the experimental setup, we shot a $0.6 \ \mathrm{mm}$ pencil beam of gammas with energies 2.6, 8, 15, and 20 MeV. Fig.~\ref{fig:gammaSimulations} plots the deposited energies in the crystal for the different beam energies. Note that no additional spectral blurring was added as a means to display the ideal spectral response of the crystal. In the plot, we observe the change in spectral profile by comparing the full energy peak to escape peak ratio. In the $2.6 \ \mathrm{MeV}$ case, the full-energy peak is the dominant peak when compared to the escape peaks. At higher energies, the escape peaks become more dominant and the products in the electromagnetic showers are more likely to escape the detector. We also observe that the largest deposited energy correlates to the incident energy.

\section{Additional Plots of Proton Data}
\label{sec:additionalPlots}

\begin{figure*}[h!]
  \centering
  \includegraphics[trim={0cm 0cm 0cm 0cm}, clip, width=0.75\textwidth]{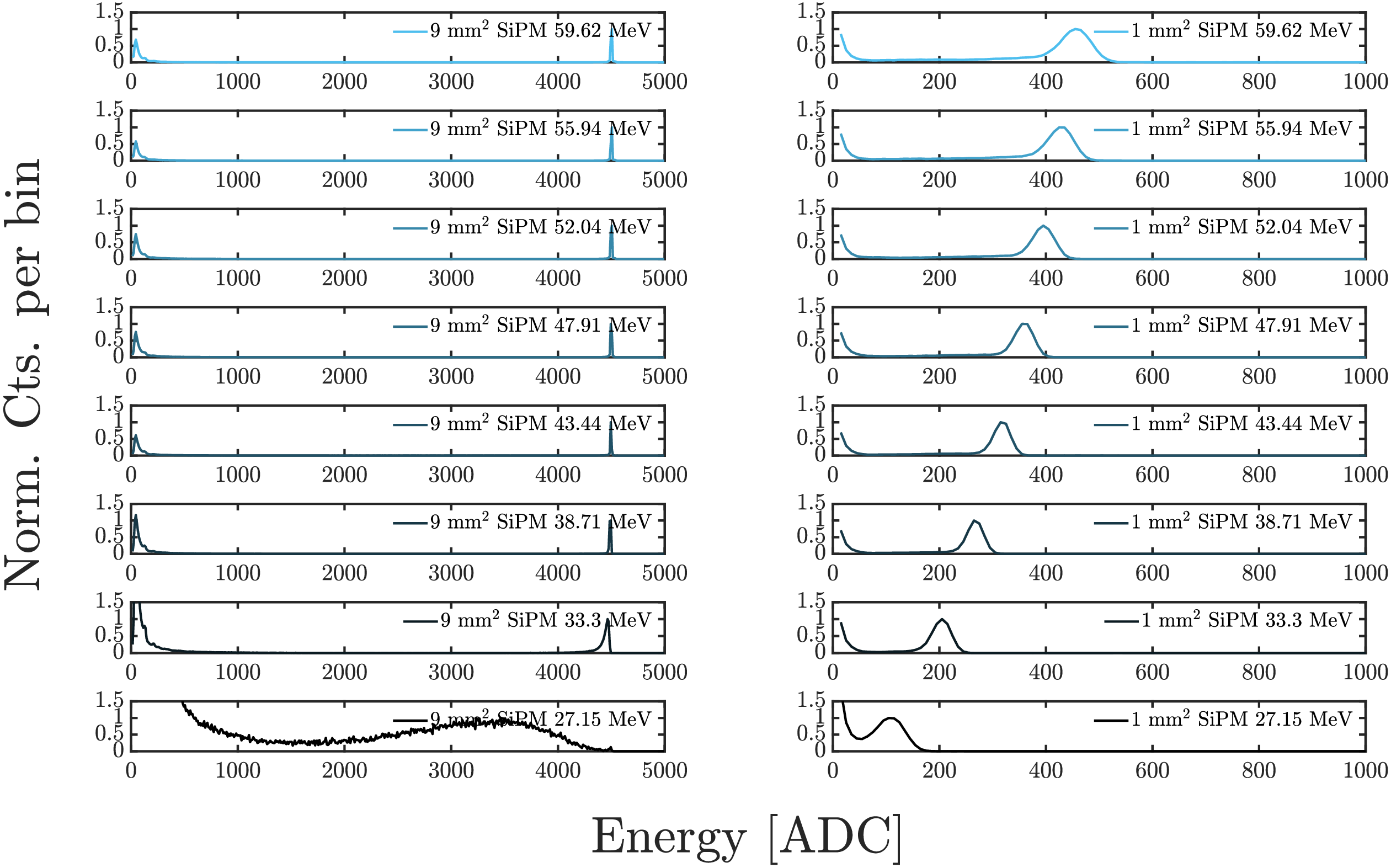}
  \caption{The response of a log to high-energy protons as measured by the V3 (left column) $9 \ \mathrm{mm}^2$ and (right column) $1 \ \mathrm{mm}^2$ SiPM. Note the overflow in left column in the plots for protons above $33.3 \ \mathrm{MeV}$ at around 4500 ADC units.}
  \label{fig:A_V3_2_low__high_gain_spectra_individual}
  
\end{figure*}

Fig.~\ref{fig:A_V3_2_low__high_gain_spectra_individual} plots the same data as Fig.~\ref{fig:A_V3_2_low__high_gain_spectra} but separates each energy to a different row. The only energy spectrum that in the SiPM range of both the $9 \ \mathrm{mm}^2$ and $1 \ \mathrm{mm}^2$ is the $27.15 \ \mathrm{MeV}$. As the energy increase from $27.15$ to $33.3 \ \mathrm{MeV}$, which is a $\sim22$\% increase in energy, would put the distribution above the dynamic range of the SiPM.

\bibliographystyle{IEEEtran}
\bibliography{IEEEbib}

\begin{thebibliography}{10}
\providecommand{\url}[1]{#1}
\csname url@samestyle\endcsname
\providecommand{\newblock}{\relax}
\providecommand{\bibinfo}[2]{#2}
\providecommand{\BIBentrySTDinterwordspacing}{\spaceskip=0pt\relax}
\providecommand{\BIBentryALTinterwordstretchfactor}{4}
\providecommand{\BIBentryALTinterwordspacing}{\spaceskip=\fontdimen2\font plus
\BIBentryALTinterwordstretchfactor\fontdimen3\font minus
  \fontdimen4\font\relax}
\providecommand{\BIBforeignlanguage}[2]{{%
\expandafter\ifx\csname l@#1\endcsname\relax
\typeout{** WARNING: IEEEtran.bst: No hyphenation pattern has been}%
\typeout{** loaded for the language `#1'. Using the pattern for}%
\typeout{** the default language instead.}%
\else
\language=\csname l@#1\endcsname
\fi
#2}}
\providecommand{\BIBdecl}{\relax}
\BIBdecl

\bibitem{mcenery2019allsky}
J.~McEnery, J.~A. Barrio, I.~Agudo, M.~Ajello, J.-M. {\'A}lvarez, S.~Ansoldi,
  S.~Anton, N.~Auricchio, J.~B. Stephen, L.~Baldini \emph{et~al.}, ``{All-sky
  medium energy gamma-ray observatory: Exploring the extreme multimessenger
  universe},'' \emph{arXiv preprint arXiv:1907.07558}, 2019.

\bibitem{COMPTEL}
V.~{Schoenfelder}, H.~{Aarts}, K.~{Bennett}, H.~{de Boer}, J.~{Clear},
  W.~{Collmar}, A.~{Connors}, A.~{Deerenberg}, R.~{Diehl}, A.~{von Dordrecht},
  J.~W. {den Herder}, W.~{Hermsen}, M.~{Kippen}, L.~{Kuiper}, G.~{Lichti},
  J.~{Lockwood}, J.~{Macri}, M.~{McConnell}, D.~{Morris}, R.~{Much}, J.~{Ryan},
  G.~{Simpson}, M.~{Snelling}, G.~{Stacy}, H.~{Steinle}, A.~{Strong}, B.~N.
  {Swanenburg}, B.~{Taylor}, C.~{de Vries}, and C.~{Winkler}, ``{{Instrument
  Description and Performance of the Imaging Gamma-Ray Telescope COMPTEL aboard
  the Compton Gamma-Ray Observatory}},'' \emph{Astrophysical Journal
  Supplement}, vol.~86, p. 657, Jun. 1993.

\bibitem{nuclearSyn}
A.~F. {Iyudin}, R.~{Diehl}, H.~{Bloemen}, W.~{Hermsen}, G.~G. {Lichti},
  D.~{Morris}, J.~{Ryan}, V.~{Schoenfelder}, H.~{Steinle}, M.~{Varendorff},
  C.~{de Vries}, and C.~{Winkler}, ``{COMPTEL observations of 44Ti gamma-ray
  line emission form CAS A},'' \emph{Astronomy and Astrophysics}, vol. 284, pp.
  L1--L4, Apr. 1994.

\bibitem{paliya2019supermassive}
V.~S. Paliya, M.~Ajello, L.~Marcotulli, J.~Tomsick, J.~S. Perkins, E.~Prandini,
  F.~D'Ammando, A.~D. Angelis, D.~Thompson, H.~Li, A.~Dominguez, V.~Beckmann,
  S.~Guiriec, Z.~Wadiasingh, P.~Coppi, J.~P. Harding, M.~Petropoulou, J.~W.
  Hewitt, R.~Ojha, A.~Marcowith, M.~Doro, D.~Castro, M.~Baring, E.~Hays,
  E.~Orlando, S.~Guiriec, V.~Bozhilov, I.~Agudo, T.~Venters, J.~McEnery, L.-S.
  The, D.~Hartmann, S.~Buson, F.~Longo, and D.~Gasparrini, ``Supermassive black
  holes at high redshifts,'' 2019.

\bibitem{burns2019opportunities}
E.~Burns, A.~Tohuvavohu, J.~M. Bellovary, E.~Blaufuss, T.~J. Brandt, S.~Buson,
  R.~Caputo, S.~B. Cenko, N.~Christensen, J.~W. Conklin, F.~D'Ammando, K.~E.~S.
  Ford, A.~Franckowiak, C.~Fryer, C.~M. Hui, K.~Holley-Bockelmann, T.~Jaffe,
  T.~Kupfer, M.~Karovska, B.~D. Metzger, J.~Racusin, B.~Rani, M.~Santander,
  J.~Tomsick, and C.~Wilson-Hodge, ``Opportunities for multimessenger astronomy
  in the 2020s,'' 2019.

\bibitem{amegox}
\BIBentryALTinterwordspacing
R.~Caputo \emph{et~al.}, ``{All-sky Medium Energy Gamma-ray Observatory
  eXplorer mission concept},'' \emph{Journal of Astronomical Telescopes,
  Instruments, and Systems}, vol.~8, no.~4, p. 044003, 2022. [Online].
  Available: \url{https://doi.org/10.1117/1.JATIS.8.4.044003}
\BIBentrySTDinterwordspacing

\bibitem{astroPix}
\BIBentryALTinterwordspacing
A.~L. Steinhebel, H.~Fleischhack, N.~Striebig, M.~Jadhav, Y.~Suda, R.~Luz,
  C.~Kierans, R.~Caputo, H.~Tajima, R.~Leys, I.~Peric, J.~Metcalfe, and J.~S.
  Perkins, ``{AstroPix: novel monolithic active pixel silicon sensors for
  future gamma-ray telescopes},'' in \emph{Space Telescopes and Instrumentation
  2022: Ultraviolet to Gamma Ray}, J.-W.~A. den Herder, S.~Nikzad, and
  K.~Nakazawa, Eds., vol. 12181, International Society for Optics and
  Photonics.\hskip 1em plus 0.5em minus 0.4em\relax SPIE, 2022, p. 121816Y.
  [Online]. Available: \url{https://doi.org/10.1117/12.2630405}
\BIBentrySTDinterwordspacing

\bibitem{compair}
\BIBentryALTinterwordspacing
D.~Shy, C.~Kierans, N.~Cannady, R.~Caputo, S.~Griffin, J.~E. Grove, E.~Hays,
  E.~Kong, N.~Kirschner, I.~Liceaga-Indart, J.~McEnery, J.~Mitchell, A.~A.
  Moiseev, L.~Parker, J.~S. Perkins, B.~Phlips, M.~Sasaki, A.~J. Schoenwald,
  C.~Sleator, J.~Smith, L.~D. Smith, S.~Wasti, R.~Woolf, E.~Wulf, and
  A.~Zajczyk, ``{Development of the ComPair gamma-ray telescope prototype},''
  in \emph{Space Telescopes and Instrumentation 2022: Ultraviolet to Gamma
  Ray}, J.-W.~A. den Herder, S.~Nikzad, and K.~Nakazawa, Eds., vol. 12181,
  International Society for Optics and Photonics.\hskip 1em plus 0.5em minus
  0.4em\relax SPIE, 2022, p. 121812G. [Online]. Available:
  \url{https://doi.org/10.1117/12.2628811}
\BIBentrySTDinterwordspacing

\bibitem{woolf2019development}
R.~S. Woolf, J.~E. Grove, B.~F. Phlips, and E.~A. Wulf, ``{Development of a
  CsI:Tl calorimeter subsystem for the All-Sky Medium-Energy Gamma-Ray
  Observatory (AMEGO)},'' in \emph{2018 IEEE Nuclear Science Symposium and
  Medical Imaging Conference Proceedings (NSS/MIC)}, 2018, pp. 1--6.

\bibitem{jSeriesArray}
\BIBentryALTinterwordspacing
``Array{J}-series - silicon photomultiplier (sipm).'' [Online]. Available:
  \url{https://www.onsemi.com/pdf/datasheet/arrayj-series-d.pdf}
\BIBentrySTDinterwordspacing

\bibitem{siPMs}
\BIBentryALTinterwordspacing
P.~Buzhan, B.~Dolgoshein, L.~Filatov, A.~Ilyin, V.~Kantzerov, V.~Kaplin,
  A.~Karakash, F.~Kayumov, S.~Klemin, E.~Popova, and S.~Smirnov, ``{Silicon
  photomultiplier and its possible applications},'' \emph{Nuclear Instruments
  and Methods in Physics Research Section A: Accelerators, Spectrometers,
  Detectors and Associated Equipment}, vol. 504, no.~1, pp. 48--52, 2003,
  proceedings of the 3rd International Conference on New Developments in
  Photodetection. [Online]. Available:
  \url{https://www.sciencedirect.com/science/article/pii/S0168900203007496}
\BIBentrySTDinterwordspacing

\bibitem{MITCHELL2022167163}
\BIBentryALTinterwordspacing
L.~Mitchell, B.~Phlips, W.~N. Johnson, M.~Johnson-Rambert, R.~Woolf, A.~Mazzi,
  and A.~Gola, ``Radiation damage assessment of sipms for scintillation
  detectors,'' \emph{Nuclear Instruments and Methods in Physics Research
  Section A: Accelerators, Spectrometers, Detectors and Associated Equipment},
  vol. 1040, p. 167163, 2022. [Online]. Available:
  \url{https://www.sciencedirect.com/science/article/pii/S0168900222005320}
\BIBentrySTDinterwordspacing

\bibitem{higs}
A.~Tonchev, M.~Boswell, C.~Howell, H.~Karwowski, J.~Kelley, W.~Tornow, and
  Y.~Wu, ``{The High Intensity $\gamma$-Ray Source (HI$\gamma$S) and Recent
  Results},'' \emph{Nuclear Instruments \& Methods in Physics Research Section
  B-beam Interactions With Materials and Atoms}, vol. 241, pp. 170--175, 2005.

\bibitem{SiPHRA}
A.~Ulyanov, D.~Murphy, A.~Fredriksen, J.~Ackermann, D.~Meier, N.~Nelms,
  B.~Shortt, S.~McBreen, and L.~Hanlon, ``{Using the SIPHRA ASIC with an SiPM
  array and scintillators for gamma spectroscopy},'' in \emph{2017 IEEE Nuclear
  Science Symposium and Medical Imaging Conference (NSS/MIC)}, 2017, pp. 1--3.

\bibitem{knoll2010radiation}
G.~F. Knoll, \emph{Radiation detection and measurement}.\hskip 1em plus 0.5em
  minus 0.4em\relax John Wiley \& Sons, 2010.

\bibitem{siPMNonlinearity}
J.~Rosado, ``Modeling the nonlinear response of silicon photomultipliers,''
  \emph{IEEE Sensors Journal}, vol.~19, no.~24, pp. 12\,031--12\,039, 2019.

\bibitem{SiPMNonlinearityOG}
\BIBentryALTinterwordspacing
D.~Renker, ``Geiger-mode avalanche photodiodes, history, properties and
  problems,'' \emph{Nuclear Instruments and Methods in Physics Research Section
  A: Accelerators, Spectrometers, Detectors and Associated Equipment}, vol.
  567, no.~1, pp. 48--56, 2006, proceedings of the 4th International Conference
  on New Developments in Photodetection. [Online]. Available:
  \url{https://www.sciencedirect.com/science/article/pii/S0168900206008680}
\BIBentrySTDinterwordspacing

\bibitem{Regazzoni_2017}
\BIBentryALTinterwordspacing
V.~Regazzoni, F.~Acerbi, G.~Cozzi, A.~Ferri, C.~Fiorini, G.~Paternoster,
  C.~Piemonte, D.~Rucatti, G.~Zappalà, N.~Zorzi, and A.~Gola,
  ``{Characterization of high density SiPM non-linearity and energy resolution
  for prompt gamma imaging applications},'' \emph{Journal of Instrumentation},
  vol.~12, no.~07, p. P07001, jul 2017. [Online]. Available:
  \url{https://dx.doi.org/10.1088/1748-0221/12/07/P07001}
\BIBentrySTDinterwordspacing

\bibitem{groveCalorimeter}
J.~E. Grove and W.~N. Johnson, ``{The calorimeter of the Fermi Large Area
  Telescope},'' in \emph{Space Telescopes and Instrumentation 2010: Ultraviolet
  to Gamma Ray}, vol. 7732.\hskip 1em plus 0.5em minus 0.4em\relax SPIE, 2010,
  pp. 138--148.

\bibitem{GECCO}
\BIBentryALTinterwordspacing
E.~Orlando, E.~Bottacini, A.~Moiseev, A.~Bodaghee, W.~Collmar, T.~Ensslin,
  I.~V. Moskalenko, M.~Negro, S.~Profumo, S.~W. Digel, D.~J. Thompson, M.~G.
  Baring, A.~Bolotnikov, N.~Cannady, G.~A. Carini, V.~Eberle, I.~A. Grenier,
  A.~K. Harding, D.~Hartmann, S.~Herrmann, M.~Kerr, R.~Krivonos, P.~Laurent,
  F.~Longo, A.~Morselli, B.~Philips, M.~Sasaki, P.~Shawhan, D.~Shy, G.~Skinner,
  L.~D. Smith, F.~W. Stecker, A.~Strong, S.~Sturner, J.~A. Tomsick,
  Z.~Wadiasingh, R.~S. Woolf, E.~Yates, K.-P. Ziock, and A.~Zoglauer,
  ``{Exploring the MeV sky with a combined coded mask and Compton telescope:
  the Galactic Explorer with a Coded aperture mask Compton telescope
  (GECCO)},'' \emph{Journal of Cosmology and Astroparticle Physics}, vol. 2022,
  no.~07, p. 036, jul 2022. [Online]. Available:
  \url{https://dx.doi.org/10.1088/1475-7516/2022/07/036}
\BIBentrySTDinterwordspacing

\bibitem{geccoSPIE}
R.~S. Woolf, A.~A. Moiseev, A.~Bolotnikov, N.~Cannady, G.~Carini, J.~F.
  Krizmanic, J.~W. Mitchell, B.~F. Phlips, M.~Sasaki, D.~Shy \emph{et~al.},
  ``{Development of the balloon-borne Galactic Explorer Coded Aperture Mask and
  Compton Telescope (GECCO) prototype},'' in \emph{Space Telescopes and
  Instrumentation 2022: Ultraviolet to Gamma Ray}, vol. 12181.\hskip 1em plus
  0.5em minus 0.4em\relax SPIE, 2022, pp. 657--669.

\bibitem{astrogam}
\BIBentryALTinterwordspacing
A.~D. Angelis \emph{et~al.}, ``{Science with e-{ASTROGAM}},'' \emph{Journal of
  High Energy Astrophysics}, vol.~19, pp. 1--106, aug 2018. [Online].
  Available: \url{https://doi.org/10.1016%2Fj.jheap.2018.07.001}
\BIBentrySTDinterwordspacing

\bibitem{cSeries}
\BIBentryALTinterwordspacing
``Micro{C}-series - silicon photomultipliers (sipm).'' [Online]. Available:
  \url{https://www.onsemi.com/download/data-sheet/pdf/microj-series-d.pdf}
\BIBentrySTDinterwordspacing

\bibitem{jSeries}
\BIBentryALTinterwordspacing
``Micro{J}-series - silicon photomultipliers (sipm).'' [Online]. Available:
  \url{https://www.onsemi.com/pdf/datasheet/microj-series-d.pdf}
\BIBentrySTDinterwordspacing

\bibitem{rosspad}
``Rosspad,'' https://ideas.no/products/rosspad/, 2022.

\bibitem{knoll2000radiation}
\BIBentryALTinterwordspacing
G.~Knoll, \emph{{Radiation Detection and Measurement}}.\hskip 1em plus 0.5em
  minus 0.4em\relax Wiley, 2000. [Online]. Available:
  \url{https://books.google.com/books?id=HKBVAAAAMAAJ}
\BIBentrySTDinterwordspacing

\bibitem{UCDavis}
C.~Castaneda, ``{Crocker Nuclear Laboratory (CNL) radiation effects measurement
  and test facility},'' in \emph{2001 IEEE Radiation Effects Data Workshop.
  NSREC 2001. Workshop Record. Held in conjunction with IEEE Nuclear and Space
  Radiation Effects Conference (Cat. No.01TH8588)}, 2001, pp. 77--81.

\bibitem{SRIM}
\BIBentryALTinterwordspacing
J.~F. Ziegler, ``{SRIM-2003},'' \emph{Nuclear Instruments and Methods in
  Physics Research Section B: Beam Interactions with Materials and Atoms}, vol.
  219-220, pp. 1027--1036, 2004, proceedings of the Sixteenth International
  Conference on Ion Beam Analysis. [Online]. Available:
  \url{https://www.sciencedirect.com/science/article/pii/S0168583X04002587}
\BIBentrySTDinterwordspacing

\bibitem{glastCalEngineering}
J.~Ampe, A.~Chekhtman, P.~Dizon, J.~Grove, W.~Johnson, B.~Leas, D.~Sandora, and
  M.~Strickman, ``{The calibration and environmental testing of the engineering
  module of GLAST CsI calorimeter},'' \emph{IEEE Transactions on Nuclear
  Science}, vol.~51, no.~5, pp. 2008--2011, 2004.

\bibitem{sipmAvalanching}
G.~Gallina, F.~Reti{\`e}re, P.~Giampa, J.~Kroeger, P.~Margetak, S.~B. Mamahit,
  A.~D.~S. Croix, F.~Edaltafar, L.~Martin, N.~Massacret, M.~Ward, and G.~Zhang,
  ``{Characterization of SiPM Avalanche Triggering Probabilities},'' \emph{IEEE
  Transactions on Electron Devices}, vol.~66, pp. 4228--4234, 2019.

\bibitem{SWORD}
E.~I. Novikova, M.~S. Strickman, C.~Gwon, B.~F. Phlips, E.~A. Wulf,
  C.~Fitzgerald, L.~S. Waters, and R.~C. Johns, ``Designing sword--software for
  optimization of radiation detectors,'' in \emph{2006 IEEE Nuclear Science
  Symposium Conference Record}, vol.~1, 2006, pp. 607--612.

\end{thebibliography}

\end{document}